\documentclass[twocolumn]{aa} % for a referee version
\usepackage{diagbox}
\usepackage{xcolor}
\usepackage{soul}
\usepackage{bm}
\usepackage[justification=centering]{caption} 
\usepackage{caption}
\usepackage{subcaption}
\usepackage{float}
\usepackage{graphicx}
\usepackage{natbib}
\usepackage{txfonts}
\usepackage{braket}
\def \be {\begin{equation}}
\def \ee {\end{equation}}
\def\restriction#1#2{\mathchoice
              {\setbox1\hbox{${\displaystyle #1}_{\scriptstyle #2}$}
              \restrictionaux{#1}{#2}}
              {\setbox1\hbox{${\textstyle #1}_{\scriptstyle #2}$}
              \restrictionaux{#1}{#2}}
              {\setbox1\hbox{${\scriptstyle #1}_{\scriptscriptstyle #2}$}
              \restrictionaux{#1}{#2}}
              {\setbox1\hbox{${\scriptscriptstyle #1}_{\scriptscriptstyle #2}$}
              \restrictionaux{#1}{#2}}}
\def\restrictionaux#1#2{{#1\,\smash{\vrule height 1.\ht1 depth 1.\dp1}}_{\,#2}}

\newcommand{\tress}{\textsc{tres}}
\newcommand{\sebaa}{\textsc{seba}}
\newcommand{\mesaa}{\textsc{mesa}}
\newcommand{\amusee}{\textsc{amuse}}

\newcommand{\tres}{\textsc{tres}\ }
\newcommand{\seba}{\textsc{seba}\ }
\newcommand{\mesa}{\textsc{mesa}\ }
\newcommand{\amuse}{\textsc{amuse}\ }
\newcommand{\sse}{\textsc{sse}\ }

\newcommand{\ms}{\,M$_\odot$}

\newcommand{\pol}{\citetalias{pol98}\ }
\defcitealias{pol98}{P98}

\newcommand{\pool}{\citetalias{pol98}}
\defcitealias{pol98}{P98}

\newcommand{\hur}{\citetalias{hur00}\ }
\defcitealias{hur00}{H00}

\newcommand{\huur}{\citetalias{hur00}}
\defcitealias{hur00}{H00}

\newcommand{\hu}{\citetalias{hur02}\ }
\defcitealias{hur02}{H02}

\newcommand{\huu}{\citetalias{hur02}}
\defcitealias{hur02}{H02}

\newcommand{\sci}{\citetalias{sci24}\ }
\defcitealias{sci24}{S24}

\newcommand{\scii}{\citetalias{sci24}}
\defcitealias{sci24}{S24}

\newcommand{\zah}{\citetalias{zah77}\ }
\defcitealias{zah77}{Z77}

\newcommand{\zahh}{\citetalias{zah77}}
\defcitealias{zah77}{Z77}

\usepackage[colorlinks=true, linkcolor=blue, citecolor=blue, urlcolor=blue]{hyperref}
\begin{document}
\title{Detailed Simulations of Massive Hierarchical Triple Star Systems}
\subtitle{Exploring the impact of the stellar physics on the evolutionary pathways of massive hierarchical triple systems}

\author{Luca Sciarini\inst{1},
Sylvia Ekstr\"om\inst{1}, Floris Kummer\inst{2}, Steven Rieder\inst{1,2,3}, Caspar Bruenech\inst{2}, Silvia Toonen\inst{2}, Eoin Farrell\inst{1}}
\institute{Department of Astronomy, University of Geneva, Chemin Pegasi 51, CH-1290 Versoix, Switzerland\\
              \email{luca.sciarini@unige.ch}
         \and
         Anton Pannekoek Institute for Astronomy, University of Amsterdam, Science Park 904, 1098 XH Amsterdam, The Netherlands
         \and
         Institute of Astronomy, KU Leuven, Celestijnenlaan 200D, 3001 Leuven, Belgium
         }

   \date{Received date ... /
Accepted date ...}
\authorrunning{Luca Sciarini et al.}
\titlerunning{Detailed Simulations of Massive Hierarchical Triple Star Systems}

% 5 {} token are mandatory
 
  \abstract
   {Recent observations estimate that approximately 30\% of early B and O-type stars are found in triple systems. So far, the evolution of triple star systems has mainly been modeled using fast stellar codes. The accuracy of these codes is expected to decrease with increasing mass, which can limit their reliability for predicting the evolutionary pathways of massive hierarchical triple star systems.}
   {We aim to investigate the discrepancies in the predicted evolution of massive stars between fast (Hurley tracks) codes and detailed (\mesaa) codes, and how these differences can impact the evolutionary pathways of massive hierarchical triple star systems.} 
   {We coupled the \tres code, which by default uses \seba to \mesa to perform the first simulations of triple systems that combine a triple secular evolutionary code with a detailed, on-the-fly stellar code. After examining the differences between the single star evolution predicted by the two stellar codes (\mesa and \sebaa), we simulate the evolution of a set of triple systems and compare their predicted evolutionary pathways under identical initial conditions.}
   {We show that among very massive stars ($M\ge50\,$M$_\odot$), the stellar tracks predicted by the two codes become increasingly divergent with increasing mass and wind mass loss efficiency. Notably, the maximal radial extent, which is crucial for determining whether the components of the triple systems interact, can differ by up to two orders of magnitude between the two stellar codes in the considered mass range and with a standard mass loss efficiency. We demonstrate that this has significant implications for the predicted evolutionary pathways of triple systems and leads to divergences between the predictions of simulations performed with \mesa and \sebaa. In particular, we show that using \mesa as stellar code, the minimum period for avoiding inner mass transfer is reduced by three orders of magnitude compared to what is predicted by \sebaa. We highlight that this result can have consequences for the formation of gravitational wave sources through the triple compact object channel.}
   {Our simulations offer new insights into the physics of triple star systems, as some relevant processes (mass loss by stellar winds, radial expansion, tides, precession due to the distortion of the stars) can be treated more self-consistently. They indicate that the results of triple systems population synthesis studies must be interpreted cautiously, in particular when the considered masses are outside the range of the grid the fast codes are based on and when significant stellar winds are considered.}


   \maketitle

%
%________________________________________________________________

\section{Introduction}
Massive stars ($M\ge8\,$M$_\odot$) are critical to our understanding of the Universe. During their short but dramatic evolution, they synthesize heavy elements, drive stellar winds and emit intense ultraviolet radiation. Their explosive deaths as supernovae release vast amounts of energy and enrich the interstellar medium of their nucleosynthetic products. These contributions make them play a pivotal role in the evolution of galaxies and the Universe \citep[e.g.,][]{woo02,cro12,hay17,eld22}.

Recent observations indicate that a large fraction of stars possess one or several companions, and that this fraction increases with stellar mass \citep[e.g.,][]{san12,san14,moe17,off23}. A significant number of these systems are likely to interact during their evolution \citep{san12}. Not only does the fraction of stars in multiple systems increase with the mass, but also does the multiplicity order (number of companions). In particular, the observed fraction of stars in triple systems among early B and O-type stars is larger than 30\% \citep{moe17}.

Although the physics of the evolution of binary systems has received extensive attention over the last two decades \citep[e.g.,][]{hur02,lan12,dem13,pax15,bel16,mar16,eld17,fra23,bav23,and24}, less is currently known about that of triple systems.

Recent studies have focused on the evolution of hierarchical triple star systems \citep[e.g.,][]{nao16a,too16,too20,ham21,ste22,pre22,too22,kum23,dor24,gen24,bru25,sha25}, where the tertiary star is situated at far enough distances from the inner binary to maintain system stability over timescales on the order of or longer than the evolutionary timescale of the stars. One of the main result of these population synthesis studies is that the presence of a third object tends to increase the rate of interactions of the inner binary as a result of the von Zeipel-Lidov-Kozai oscillations \citep[ZLK,][]{von10,lid62,koz62}. Notably, \citet{kum23} proposed statistical predictions of the main evolutionary pathways of massive hierarchical triple star systems following a population synthesis approach.

The large majority of the triple systems simulations studies have been performed utilizing rapid stellar codes based on the fitting formulae by \citep[][hereafter \huur, \huu]{hur00,hur02}. This family of fast stellar codes notably include \sse \citep{hur00}, \seba \citep{por96,too12}, \textsc{cosmic} \citep{bre20}, \textsc{compass} \citep{ril22}, \textsc{StarTrack} \citep{bel02} . One exception is the study by \citealt{por16} where they coupled a detailed stellar code to a $N$--body code to perform simulations of hierarchical triple systems. Despite their outstanding computing performances, fast stellar codes suffer from simplifying assumptions and some processes are not treated self-consistently (e.g. wind mass loss, mass transfer, ...). The limitations of the fast stellar codes have been known since their conception, but less well studied is the impact of these simplifications on the evolution of multiple systems. \citet{bav23} showed that by not self-consistently modeling the reaction of the star to mass loss by stellar winds, fast stellar codes can in some cases overpredict the maximal radial expansion of massive stars by several orders of magnitude, which can have dramatic consequences for the predicted evolution of binary systems. The impact of the discrepancies between stellar tracks predicted by detailed and rapid evolutionary codes on the evolution of triple systems has received little attention.

Although there have recently been new approaches and alternatives to the \hur formulae, which have proven successful in overcoming some of their limitations \citep[e.g.,][]{spe15,eld17,kru18,agr23,fra23}, these methods have not yet been applied in triple systems population synthesis studies, except in the very recent work by \citep{sha25b}. In this study, they coupled a triple secular code to the \textsc{posydon} tracks \citep{fra23} to investigate the origin of the V404 low-mass X-ray binary. They report differences between \textsc{posydon} and \sse stellar evolution tracks and discuss the impact of these differences on their triple systems simulations.

In this study, we coupled the TRiple Evolution Simulation code \citep[\tress,][]{too16} to the Modules for Experiment in Stellar Astrophysics \citep[\mesaa,][]{pax11,pax13,pax15,pax18,pax19} to perform the first simulations of triple systems combining a triple secular evolutionary code with a detailed, on-the-fly stellar code. Our approach differs from that of \citep{sha25b} as in our simulations the detailed stellar evolution is computed simultaneously to the three-body dynamics. This allows for alterations of the stellar physics on a per system basis, and provides access to the full structure of the stellar components, enabling the self-consistent modeling of key interactions specific to triple systems (in particular the precession caused by the distortion of the stellar components and its competition with the precession induced by the ZLK mechanism). 

We primarily focused on stars with masses $M\ge 50$\,M$_\odot$, as they are frequently found in triple configurations and because above this mass threshold the \hur tracks extrapolate the fitting formulae obtained from the \citep[][hereafter \pool]{pol98} grid.
We followed an approach similar to that in \citep{bav23}, aiming to illustrate the importance of a more self-consistent stellar evolution modeling for predicting the evolutionary pathways of massive triple systems. To this end, we compared the outcome of our simulations to analogous simulations performed with a fast stellar code \citep[\sebaa,][]{por96,too12}.

The structure of this paper is as follows: in Sect. \ref{methods}, we provide a description of the methods we used and the assumptions of single and triple star physics we made. In Sect. \ref{single_star}, we compare the single star evolution provided by the two different stellar codes we used. In Sect. \ref{triple_systems}, we present simulations of a set of triple systems for which the two codes predict divergent evolutionary pathways. Finally, we summarize our findings in Sect. \ref{conclusion}.
\section{Methods}\label{methods}
To investigate the evolutionary pathways of massive hierarchical triple systems, we used the TRiple Evolution Simulation code \citep[\tress,][]{too16}. \tres couples a stellar evolution code to a secular evolution code \citep[\textsc{seculartriple}\normalsize,][]{too16} that solves the 3--body dynamics of the system following the secular approximation. In this approach, the perturbations occasioned by the third body are assumed to act on timescales much longer than the dynamical timescale of the inner binary, so that only the long-term (secular) evolution is considered, following a perturbative method. By default, the stellar evolution in \tres is handled by \sebaa, a rapid population synthesis code based on the fitting formulae of \hur.

\tres makes use of the Astrophysical MUltipurpose Software Environment \citep[\amusee,][]{por09,pel13,por13}, a library of codes that allows the user to easily exchange between codes according to the necessity of the modeling.

\mesa \citep{pax11,pax13,pax15,pax18,pax19} is a 1d detailed stellar evolution code, which is extensively used in the field of single star and binary systems modeling. \mesa is interfaced in \amuse in its version r15140. Utilizing the \mesa interface in \amusee, we coupled \mesa to \tress, which allowed us to perform state-of-the-art, detailed simulations of hierarchical triple systems. The code developments required to make \tres fully compatible with \mesa have been made publicly available\footnote{\tres (including the developments to make it compatible with \mesaa) is available on: \url{https://github.com/ amusecode/TRES}, the \amuse main repository is accessible on: \url{https://github.com/amusecode/amuse}.}.

The approach chosen in this study is to use the versatility of \amuse to compare the evolutionary outcome of triple systems obtained with a rapid code (\sebaa)  to that obtained with a detailed code (\mesaa). For all the presented systems, the evolution was computed both with \seba and \mesaa, keeping the exact same physics for the binary and triple interactions. 
\subsection{Connecting TRES and MESA}\label{tresmesa}
The connection between \tres and \mesa was done through the interface of \mesa in \amusee. In \amusee, \mesa is treated as a library. Parameters are set with a \texttt{set\_control} method, instead of the usual \texttt{inlist} file. Additional getters and setters functions are available, which allow communications with the \mesa instances.

Some binary and triple interactions implemented in \tres depend on quantities directly related to the structure of the stars, such as the apsidal motion constant $k_{\rm AMC}$ and the gyration radius $r_{\rm g}$ (see Sect. \ref{triple_physics} for more details). When using \tres with \sebaa, these quantities are given rough values which depend mainly on the evolutionary phase of the star. For instance, the apsidal motion constant is assumed to be $k_{\rm AMC}=0.0144$ for main sequence (MS) stars, helium-MS stars, and white dwarfs, $k_{\rm AMC} = 0.260$ for neutron stars, $k_{\rm AMC} = 0$ for black holes (BHs) and $k_{\rm AMC} = 0.05$ for all other types of stars, following the prescriptions of \citep{bro55,cla92}\footnote{\seba also incorporates an implementation of a more detailed modeling of $k_{\rm AMC}$, but it is only valid for masses up to $\sim 10 M_\odot$.}. In reality, $k_{\rm AMC}$ and $r_{\rm g}$ vary throughout the evolution of the stars \citep[e.g.,][]{cla89,cla99,cla10,ros20}, see also Appendix \ref{AppC}. When using \tres with \mesaa, these quantities are directly retrieved from the structure, which offers a more self-consistent treatment of the corresponding interactions. For that purpose, we added the methods \texttt{get\_apsidal\_motion\_constant} and \texttt{get\_gyration\_radius} to the \mesa interface in \amusee.

In \tress, the binary and triple interactions are handled by \textsc{seculartriple}. For this reason, we used the single star module of \mesa instead of the binary module. All the interactions are thus handled by \tress. As in \citep{kum23}, we did not follow in detail the mass transfer (MT) events but stopped the simulations whenever a star experiences Roche Lobe overflow (RLOF). Therefore in the presented simulations, the stars' evolution affect the orbital evolution of the system, but the feedback from the interactions to the stellar evolution are not accounted for. As a result, the stellar evolution is identical to that of isolated single stars. The list of stopping conditions are given in Sect. \ref{triple_physics}, together with a description of the binary and triple interactions.
\subsection{Single star physics}\label{stellar_ingredients}
For the comparisons of the evolutionary pathways obtained with \mesa and \seba to be pertinent, it is crucial to use the two codes with equivalent stellar physics ingredients. We sought to follow this approach as precisely as possible. Some stellar physics ingredients however could not be kept exactly consistent between the two codes, due to intrinsic differences between rapid and detailed codes. In this section we discuss our choice of input physics and disparities between the two stellar codes.

Rapid codes such as \seba are based on the formulae of \hur fitted to the tracks of \pol. The \pol grid covers the mass range $M\in [0.5,\, 50]$\,M$_\odot$. Although nothing prevents the use of the fitting formulae for stars outside this range of mass, they get increasingly divergent from tracks computed with a detailed code for increasing masses, as illustrated in Sect. \ref{grid_comparison}.

The grid of \pol was computed without accounting for mass loss by stellar winds. In their paper, \hur provide receipes for incorporating mass loss a posteriori in rapid codes. In their approach, after a star has lost a certain amount of mass due to stellar winds, it shifts to the track of a star of its current mass. The reaction to mass loss is therefore not self-consistent. In our simulations, we sought to use equivalent winds prescriptions with the two codes, and show that an inconsistent treatment of the reaction to mass loss can imply divergent evolutionary pathways for certain triple systems. The reactions to mass loss obtained with the two codes are compared in  Sect. \ref{mass_loss_impact}.
\subsubsection{Metallicity}
We simulated systems at solar metallicity. The grid of \pol was computed with the old value $Z_\odot=0.02$, which we adopted for all simulations. This is also the default value of the solar metallicity in \mesa (in particular for the winds scaling, see Sect. 2.2.3 in \citealt{and24}), so no adaptations were needed to use $Z_\odot=0.02$.
\subsubsection{Rotation}
The \pol grid does not account for the effect of rotation on stellar evolution. This is the main reason why we decided to only compute non-rotating \mesa models. The second reason is related to the coupling of the stellar rotation to the overall evolution of the system and is discussed in Sect. \ref{orbital_evolution}
\subsubsection{Convection}
In our \mesa simulations, we adopted convection physics as close as possible to that of the \pol grid. We used the same value of $\alpha_{\rm MLT}=2.0$. We also did not account for semi-convection. In order to reduce the superadiabaticity in radiation-dominated convective regions of the most massive stars, we used MLT++ as in \citep{cho16}. 

In the \pol grid, they used a modified criterion for convective stability that is not available in the default \mesa version. In their modeling, a layer is convective whenever $\nabla_{\rm rad} \ge \nabla_{\rm ad}+ \delta$, where $\delta$ is the product of a free parameter $\delta_{\rm ov}$, their overshooting constant, and a factor which depends on the ratio of radiation to gas pressure $\zeta$. The effect of this prescription corresponds to a step overshoot with a decreasing value of $\alpha_{\rm ov}$ along the MS\footnote{In their modeling, $\delta$ decreases when $\zeta$ increases along the MS.}.

Instead of implementing the \pol prescription in \mesaa, we chose to follow a more standard approach by using a step overshoot prescription for the \mesa simulations. We calibrated the value of $\alpha_{\rm ov}$ as to fit the maximum radius reached during the MS by the 50\,M$_\odot$ \seba star, which corresponds to the upper mass limit of the \pol grid. As will be discussed in Sect. \ref{mass_loss_impact}, the reaction to mass loss of the \hur tracks is not self-consistent and alters the maximum radius reached during the MS. In order to free ourselves from this issue, we considered necessary to perform the $\alpha_{\rm ov}$ calibration without accounting for any mass loss, as the tracks are reliable in this case (at least for the MS evolution and for a mass inside the range of the \pol grid). As to illustrate our point, we show in Sect. \ref{mass_loss_impact} a comparison of the maximum MS radius reached with \mesa and \seba for different initial masses and mass loss scaling factors.

Following this approach, we obtained the value $\alpha_{\rm ov}=0.31$, which we adopted for all masses. We note that this value is similar to that used in the \citep{bro11} models (they used $\alpha_{\rm ov}=0.335$). As \pol show in their Fig. 1, their prescription corresponds to an increasing Zero Age Main Sequence (ZAMS) value of $\alpha_{\rm ov}$ with the mass. This calibrated value may therefore be overestimated for the lower mass stars and underestimated for the higher mass stars. However, in the mass range we considered ($M\in[8,120]\,$M$_\odot$), the corresponding ZAMS value of $\alpha_{\rm ov}$ stemming from \pol convection prescription only moderately increases with the mass, ranging from 0.3 to 0.42.

The convection criteria in \pol implies a decreasing value of $\alpha_{\rm ov}$ along the MS. In order to mimic this treatment, we computed a \mesa toy model starting with a value of $\alpha_{\rm ov}$ higher than the calibrated one and manually reducing it along the MS. We did not find significant differences between the MS evolution of this model and the one with a constant value of $\alpha_{\rm ov}=0.31$ in terms of radial expansion (see Appendix \ref{AppA} for more details).

Based on these considerations, we judge that the difference in the treatment of overshoot between the two codes does not affect significantly our results.
\subsubsection{Stellar winds}
In our \mesa simulations, we used the \texttt{Dutch} wind combination as in \citep{gle09,cho16}. It consists of the \citep{vin01} prescription for the temperature range 10'$000\,\text{K} \le T_{\rm eff}\le  50$'000\,K, the \citep{dej88} for $T_{\rm eff}\le 10$'000\,K and the \citep{nug00} Wolf-Rayet (WR) prescription for stars with a hydrogen surface mass fraction $X<0.4$ and $T_{\rm eff}\ge$ 10'000\,K. Observations by \citep{bou05,ful06} and simulations by \citep{sun10} indicate that the usual prescriptions in general overestimate the mass loss rates by a factor 2 or more. For this reason, we lowered the prescriptions by a factor 3 using \texttt{Dutch\_scaling\_factor=0.333} as done in \citep{kum23}. We recall that wind mass loss rates in the range of masses we consider are still subject to large uncertainties \citep[e.g.,][]{vin01,kri21,bra22,vin22,bjo23}. For specific systems whose evolution is strongly affected by mass loss, we investigated the choice of the wind prescription by comparing the obtained evolutionary pathways according to the wind scaling factor and the chosen stellar code.

In the \seba simulations, we used the same set of prescriptions except for a small difference in the WR winds prescription. The \citep{nug00} prescription in principle requires a value for the surface hydrogen mass fraction, and is metallicity dependent. More precisely, the \citep{nug00} prescription usually used in detailed stellar models (their equation (22)) depends on the luminosity $L$, helium mass fraction $Y$ and metallicity $Z$ as:
\begin{equation}
    \dot M = 10^{-11}\left(\frac{L}{\text{L}_\odot}\right)^{1.29} Y^{1.7}Z^{0.5},
    \label{wr_metal}
\end{equation}
and is applied when $X<0.4$. Rapid codes such as \seba do not follow in detail the evolution of the abundance profile of the star, and therefore do not provide the values of $X,\ Y, \ Z$, which are necessary to use this prescription. In order to get around this issue, we decided to use Eq. (25) in \citep{nug00} for the \seba simulations. In their paper, they provide different fits to the observed mass loss rates of WR stars. Equation (25) in \citep{nug00} reads:
\begin{equation}
    \dot M = 10^{-5.73}\left(\frac{M}{\text{M}_\odot}\right)^{0.88},
    \label{wr_m}
\end{equation}
and depends only on the mass, which makes it usable for the \seba simulations. Because the WR mass loss rates are in principle metallicity dependent, \citet{nug00} recommend to use Eq. \eqref{wr_metal}, but given the nature of \sebaa, we consider that using Eq. \eqref{wr_m} provides an acceptable alternative.

The WR winds treatment between \mesa and \seba thus differ in two aspects. The first is that two different formulas are used. The second is in what activates the wind prescription. In the \mesa simulations, the WR winds are activated when the surface hydrogen mass fraction becomes smaller than 0.4. In contrast, in the \seba simulations, they occur when the star becomes a helium star, i.e. when it has lost its whole envelope\footnote{The transition to the helium star phase occurs when the star has lost its whole envelope, of mass $M_{\rm env}=M_{\rm tot}-M_{\rm core}$, where the core mass is obtained with different equations in \hur according to the evolutionary stage. In particular, the core mass is set to 0 during the whole MS, which implies that following the \hur fitting formulae models never become helium stars before the end of the MS.}. As will be discussed in Sect. \ref{mass_loss_impact}, these two different approaches can in some cases lead to significantly different evolution in the Hertzsprung-Russel diagram (HRD).

In Sect. \ref{grid_comparison}, we compare the \mesa and \seba single star MS evolution with our default wind prescription (\texttt{Dutch} combination lowered by a factor 3). We then investigate the impacts of mass loss by comparing the evolution with stronger winds (standard \texttt{Dutch} combination) in Sect. \ref{mass_loss_impact}.
\subsection{Triple systems physics}\label{triple_physics}
The physics of the triple system adopted in this study is identical to that in \citep{too16,kum23} (apart from the destabilization criterion, see below). We recall here the relevant components.
\subsubsection{Stability of the triple system}
To ensure stability over a long timescale, the system needs to be hierarchical, in the sense that the outer separation $a_{\rm out}$ needs to be somewhat larger than the inner separation $a_{\rm in}$. In such configuration, the timescale of the perturbation implied by the tertiary component is large compared to the dynamical timescale of the inner binary. When the orbit of the tertiary becomes too close to that of the inner binary, this is not the case anymore, and the secular approximation breaks down. A proper 3--body dynamics integration is required to follow the chaotic behavior of the orbits. As shown in \citep[e.g.,][]{ham22, too22,bru25}, the system may in this case dissolve into either a single star and a binary or three single stars, be subject to a collision, or preserve its hierarchy. By default, in \tres the stability criterion is that of \citep{mar99,mar01}. In this study, we used the more recent criterion of \citep{vyn22}, based on $N$--body simulations. In their paper, they provide a publicly available python routine that follows a machine learning approach, which performs better than the criterion by \citep{mar01} for determining the stability of triple systems. They also provide a fitting formula to their result, which reads:
\begin{equation}
\begin{split}
\restriction{\frac{a_{\rm out}}{a_{\rm in}}}{\rm crit}=&\ 2.4\frac{1+\widetilde{e}_{\rm in}}{1-e_{\rm out}}\left(\frac{1+q_{\rm out}}{(1+\widetilde{e}_{\rm in})\sqrt{1-e_{\rm out}}}\right)^{2/5}\\&\times\left[\frac{1-0.2\widetilde{e}_{\rm in}+e_{\rm out}}{8}\left(\cos i_{\rm mut} - 1 \right)+1\right],
\end{split}
\label{dyn_des}
\end{equation}
where $e_{\rm out}$ is the outer eccentricity, $i_{\rm mut}$ the mutual inclination, $q_{\rm out}=\frac{M_3}{M_1+M_2}$ where $M_1,\ M_2$ and $M_3$ are the masses of the components. The term $\widetilde{e}_{\rm in}$ is defined in Eq. (3) of \citep{vyn22}. In our simulations we used Eq. \eqref{dyn_des} as the definition of the critical ratio of $\frac{a_{\rm out}}{a_{\rm in}}$. Systems are stable when:
\begin{equation}
\frac{a_{\rm out}}{a_{\rm in}}>\restriction{\frac{a_{\rm out}}{a_{\rm in}}}{\rm crit}.
\label{criterion}
\end{equation}
Eq. \eqref{criterion} defines a parameter space of stable orbits. In all the presented simulations, the systems are born stable. Some of them may destabilize at some point of their evolution. This is typically the case when the inner binary is subject to significant mass loss, which makes $a_{\rm in}$ increase and $q_{\rm out}$ decrease. The modeling of triple systems in the dynamical unstable regime with a detailed stellar code was beyond the scope of this project. We therefore stop our simulations whenever criterion \eqref{criterion} is violated.
\subsubsection{3--body dynamics and ZLK oscillations}
The lowest order (i.e. quadrupole order) perturbation of the third component to the orbital evolution of the inner binary is the ZLK mechanism. It consists of an exchange of angular momentum between the inner and the outer orbit. In some cases, it can induce significant oscillations of the eccentricity of the inner orbit and the mutual inclination. 

In the approximation of the test-particle, and for an initially circular orbit, the maximum inner eccentricity reached through ZLK oscillation only depends on the initial mutual inclination:
\begin{equation}
    e_{\max}=\sqrt{1-\frac{5}{3}\cos^2(i_{\rm mut})}.
\end{equation}
In this approximation, ZLK cycles only occur when the initial inclination is in the range $i_{\rm mut}\in[39.2$º$,\ 140.8$º$]$ \citep{for00,nao16}.

For systems with larger values of $a_{\rm in}/a_{\rm out}$, i.e. with a less pronounced hierarchy, additional terms in the secular approximations are needed. The next (octupole) term in the secular approximation can lead to extreme eccentricity oscillations for systems with a non-zero outer eccentricity and an inner binary with unequal stellar masses. The importance of the octupole term with respect to the quadrupole term can be measured with the octupole parameter, defined by:
\begin{equation}
    \epsilon_{\rm oct}=\frac{M_1-M_2}{M_1+M_2}\frac{a_{\rm in}}{a_{\rm out}}\frac{e_{\rm out}}{1-e_{\rm out}^2}.
    \label{octupole}
\end{equation}
Systems with an octupole parameter $|\epsilon_{\rm oct}|\gtrsim 0.001$ are said to be in the "eccentric ZLK regime", where the octupole term can play a significant role \citep[e.g.,][]{lit11,sha13}.
\subsubsection{Precession}\label{precession}
Binary and triple systems can be subject to precession of the orbit. In the case of binary systems, precession can affect the orientation of the orbit, but not the overall evolution of the system. 

In triple systems, the picture is more complicated as ZLK cycles also induce a precession of the orbit. Other causes of precession (distortion of the star due to its intrinsic rotation or tides, general relativistic effects) can enter in competition with the precession driven by the ZLK oscillations and quench it, which can in turn alter the evolution of other orbital parameters (eccentricity, inclination, ...) (see \citealt{nao16} for a comprehensive review of the ZLK mechanism).
The different sources of precession included in \tres are listed in \citep{too16}. Among them, we recall the equation driving the precession caused by the tidal distortion of the star \citep{sme01}:
\begin{equation}
    \dot g_{\rm tides}=15\frac{k_{\rm AMC}}{\left(1-e^2\right)^5}\Omega_{\rm orb}\left(1+\frac{3}{2}e^2+\frac{1}{8}e^4\right)\frac{M_2}{M_1}\left(\frac{R}{a}\right)^5,
    \label{prec_tid}
\end{equation}
and by intrinsic stellar rotation \citep{fab07}:
\begin{equation}
    \dot g_{\rm rotate}=\frac{k_{\rm AMC}}{\left(1-e^2\right)^2}\frac{\Omega_{\rm spin}^2}{\Omega_{\rm orb}}\frac{M_1+M_2}{M_1}\left(\frac{R}{a}\right)^5.
    \label{prec_rot}
\end{equation}
In Eq. \eqref{prec_tid} and \eqref{prec_rot}, $\Omega_{\rm orb}$ stands for the orbital angular velocity of the binary system, $\Omega_{\rm spin}$ the angular velocity of the star, $R$ its radius. These two equations depend on the apsidal motion constant of the star, which can be directly determined from the structure of the stars in the \mesa simulations, offering a more self-consistent modeling of the precession of the triple system, and hence of its overall orbital evolution (see Sect. \ref{orbital_evolution}). 
\subsubsection{Tidal interactions}
Unless stated otherwise, the chosen tides formalism in the presented simulations is that of \citep[][hereafter \huu]{hur02}. We recall that under this formalism, both the equilibrium and dynamical tides are incorporated into the formalism of \citep{hut81}. 

In \citep[][hereafter \scii]{sci24}, we demonstrated that following this procedure, the tidal evolution obtained in the case of the dynamical tides is not consistent with the original formalism by \citep[][hereafter \zahh]{zah77}. In this letter, we obtained an updated set of equations consistent with the original formulation by \zahh. Yet, for this study we chose in most cases to stick to the \hu formalism. There are mainly two reasons for this choice. 

The first is that the approach followed in this study is to primarily investigate the effect of the detailed stellar physics onto the evolution of the triple system, keeping the interactions unchanged. The present study complements that of \citep{kum23}, who investigated the evolutionary pathways of massive hierarchical triple systems. We decided to keep the same physics for the interactions as in their study.

The second is that the derivation proposed by \zahh, which we complemented in \scii, is an expansion keeping only low-order eccentricity terms. It is well suited for systems with moderate eccentricity (say $e\lesssim 0.5)$, but likely to be less appropriate for systems with large eccentricities, which are often found among triple and higher order systems due to the ZLK oscillations. The equilibrium tides formalism of \citep{hut81} assumes that tides make the star reach an equilibrium shape with a constant time lag between the tidal bulge and the line of centers of the two bodies, an assumption not well justified in the case of the dynamical tides where the star oscillates. Its strength lies in its applicability to systems with any orbital eccentricity. Under the formalism of \huu, the functional dependence of the eccentricity and the angular velocity in the case of the dynamical tides is exactly the same as in the case of the equilibrium tides formalism of \citep{hut81}. It might be more suitable for systems with very high eccentricities, even though its functional dependence (in particular the angular velocity dependency) intrinsically comes from the formalism of \citep{hut81}, i.e. a formalism of equilibrium tides. 

In Sect. \ref{ZLKTF}, we compare the evolution obtained with the \hu formalism to that obtained with the \sci prescription (updated \zahh) for a couple systems subject to ZLK oscillations and tidal interactions.

\subsubsection{Mass transfer}
Because triple systems are subject to ZLK oscillations, mass transfer frequently occurs in eccentric orbits \citep{too20}. Mass transfer in eccentric orbits is still poorly understood \citep[see e.g.,][]{sep07,dav13,ham19}. Whenever a star fills its Roche Lobe, the simulation is stopped and mass transfer (either in the inner binary or from the tertiary star) is considered as the evolutionary pathway of the system. It is therefore appropriate to use the value of the Roche Lobe radius at periastron to determine if a star fills its Roche Lobe. In this case, the expression of \citep{egg83} needs to be corrected with an eccentricity dependent term:
\begin{equation}
    R_{{\rm L},i}= a_j\frac{0.49q_i^{2/3}}{0.6q_i^{2/3}+\ln\left(1+q_i^{1/3}\right)}\left(1-e_j\right),
\label{roche_lobe}
\end{equation}
where $q_1=\frac{M_1}{M_2}$, $q_2=\frac{M_2}{M_1}$, $q_3=\frac{M_3}{M_1+M_2}$, $j$ refers to the inner ($i=1,2)$ or outer orbit $(i=3)$. As such, for the inner binary Eq. \eqref{roche_lobe} neglects distortions to the Roche geometry caused by the third star.
\subsubsection{Orbital evolution and coupling to stellar evolution}\label{orbital_evolution}
The set of equations governing the orbital evolution of the triple system is given in \citep{too16}. In \tress, it is solved within the \textsc{seculatriple} code. Notably, the evolution of the stars' angular velocity is included in the set. The rotation of the star is thus treated outside the stellar evolution code. The main reason is that by default \tres uses \seba as stellar evolution code, which does not account for the effect of rotation on stellar evolution.

During a simulation with \tress, the stellar and secular evolution codes are called successively until the next computed age (\texttt{next\_age}). In order to reach the \texttt{next\_age}, different timesteps are generally used by the stellar and the secular evolution codes. The main reason is that depending on the orbital configurations, the secular timescale can differ significantly from the timescale of stellar evolution. The angular velocities being included inside the set of equations governing the orbital evolution, their evolution is solved for together with the quantities whose evolution is altered by the dynamics of the system. For this reason, the consistent coupling of the rotation of the star to the overall evolution of the system is not straightforward. In order to account for the effect of rotation onto stellar evolution, stellar and orbital evolution would need to share the same timesteps, which might be very constraining for the stellar evolution when the ZLK timescales are short with respect to the stellar evolution timescales.

For these reasons, we followed a simpler approach by neglecting the effects of stellar rotation on stellar evolution. This is the second reason why there are no feedback from the interactions to the stellar evolution in our simulations.
\subsubsection{Evolutionary pathways}\label{evolutionary_path}
The possible evolutionary pathways of the triple systems in this study are the same as in \citep{kum23}. They consist of stopping conditions of the presented simulations:
\begin{enumerate}
    \item Mass transfer in the inner binary when the primary or the secondary fills its Roche Lobe.
    \item Mass transfer from the tertiary to the inner binary when the tertiary fills its Roche Lobe.
    \item Dynamical destabilization when criterion \eqref{criterion} is violated.
    \item Dynamical unbinding of one of the orbit as a result of core collapse supernova of one of the three components.
    \item None of the above interactions occur within a Hubble time. In this case the systems are considered as non-interacting.
\end{enumerate}

\section{Single star evolution}\label{single_star}
One of the main driver of the evolution of massive hierarchical triple systems is the stellar evolution of its components. The final state of the system depends on the interactions the stars experience during their lives, which are strongly impacted by their radial expansions and mass lost by stellar winds. In our simulations we do not account for feedback from the interactions onto stellar evolution. Thus, it is sufficient to get a satisfactory understanding of the differences in the single star evolution provided by \mesa and \seba to understand how they can imply divergences in the predicted evolutionary pathways of triple systems. In this section we aim at comparing and explaining the obtained differences in the single star evolution provided by \mesa and \seba under the choice of stellar physics ingredients presented in Sect. \ref{stellar_ingredients}.
\subsection{Grid comparison}\label{grid_comparison}
\begin{figure*}[h]
\begin{subfigure}[b]{0.5\textwidth}
\centering
\centerline{\includegraphics[trim=0.5cm 0cm 1.8cm 1.5cm, clip=true, width=.99\columnwidth,angle=0]{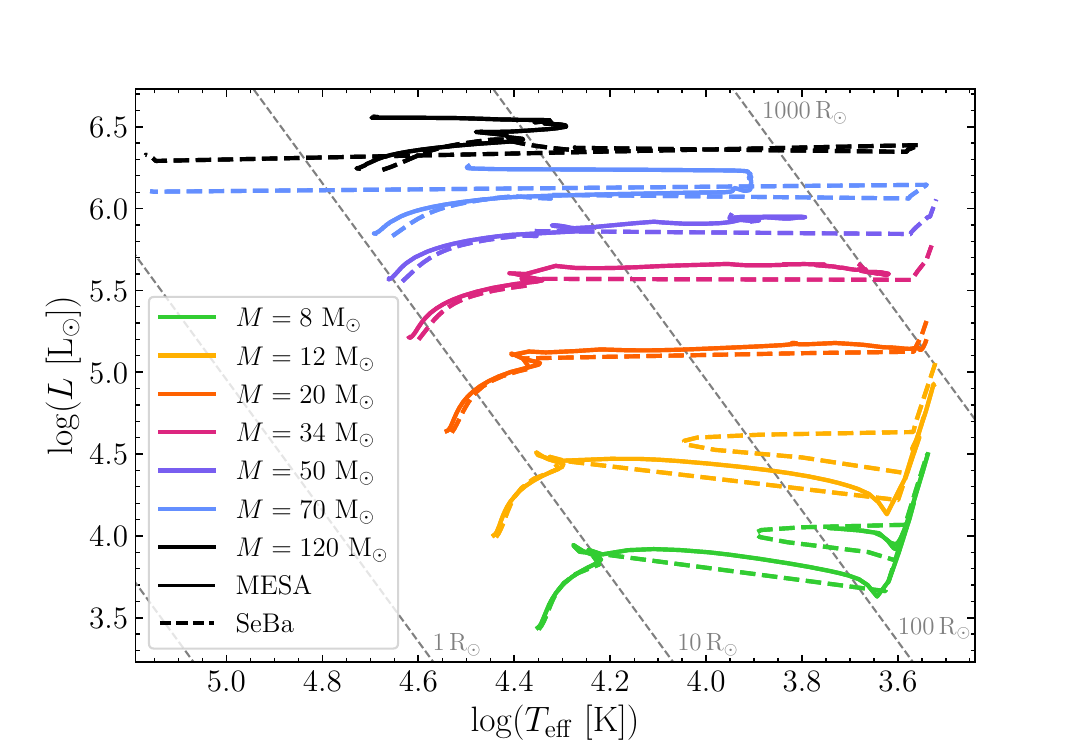}}
\caption{
HRD for stars in the mass range $M\in[8,\, 120]\,$M$_\odot$.}
\label{grid_comp}
\end{subfigure}
\hfill
\begin{subfigure}[b]{0.5\textwidth}
\centering
\centerline{\includegraphics[trim=0.5cm 0.01cm 1.88cm 1.4cm, clip=true, width=1.01\columnwidth,angle=0]{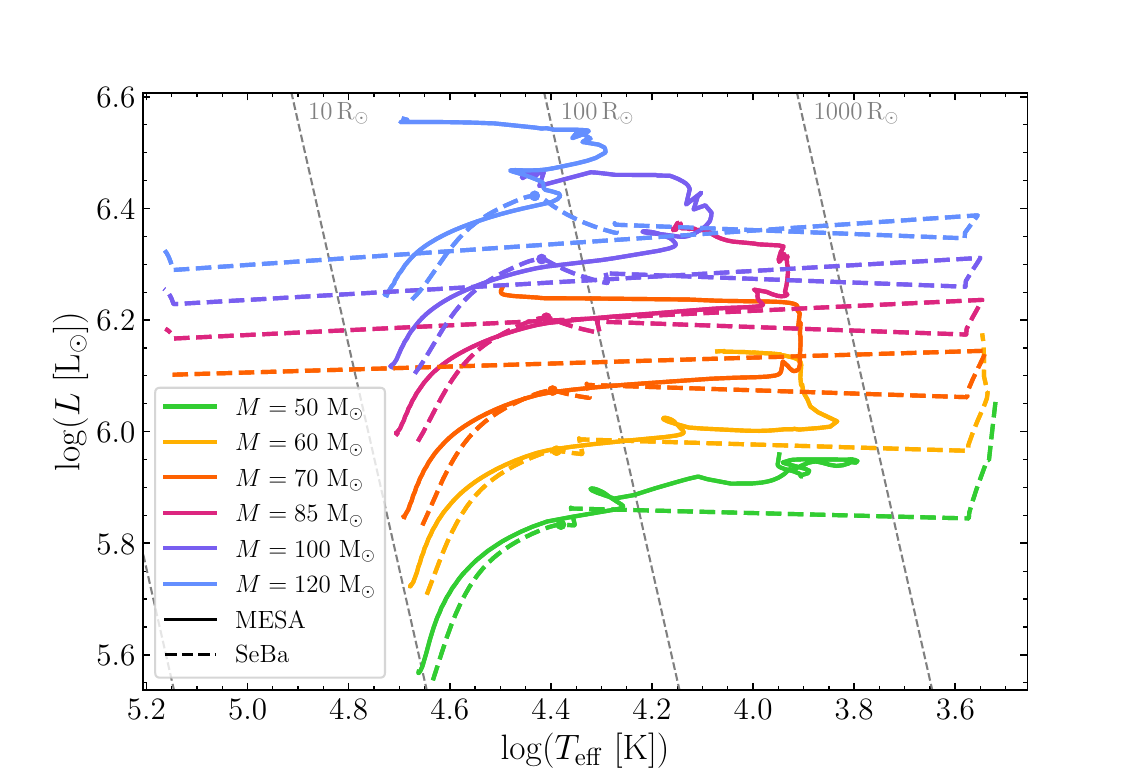}}
\caption{
HRD for stars in the mass range $M\in[50,\, 120]\,$M$_\odot$.}
\label{HRD_massive}
\end{subfigure}
\hfill
\caption{HRD obtained with \mesa (solid lines) and \seba (dashed lines). Isoradius lines are represented by gray dashes.}
\label{grid_comp_both}
\end{figure*}

The HRD of massive stars with initial masses in the range $M\in[8,\, 120]\,$M$_\odot$ obtained with \mesa and \seba and the stellar physics ingredients described in Sect. \ref{stellar_ingredients} is shown in Fig. \ref{grid_comp}.
Comparing \mesa and \seba tracks, we observe that:
\begin{enumerate}
    \item The chosen calibration of $\alpha_{\rm ov}$ offers a satisfactory match between the MS width of \mesa and \seba models among the less massive stars. Indeed, the location of the MS hook for stars in the mass range $M\in [8,\,50]\,$M$_\odot$ (i.e. for stars inside the range of the \pol grid) is relatively consistent.
    \item The late evolution of \mesa and \seba models is relatively consistent in the mass range $M\in [8,50]\,$M$_\odot$, although the maximum radius reached during the evolution is slightly higher with \sebaa.
    \item \mesa and \seba tracks exhibit increasing discrepancies with mass, especially for $M\ge 50$\,M$_\odot$, i.e. outside the range of the \pol grid. In particular, the maximum radius differs by about an order of magnitude in this mass range.
\end{enumerate}
In order to get a closer look at the most massive stars (outside the range of the \pol grid), we show in Fig. \ref{HRD_massive} a denser HRD for stars in the range $M\in[50,\,120]\,$M$_\odot$, which allows to observe the following differences:
\begin{enumerate}
    \item The \mesa MS width in this range of mass and with this choice of wind scaling factor is larger than that of \sebaa. \mesa stars expand more during their MS, therefore the \mesa MS width is larger among massive stars. The simplest explanation for this difference is that stars of these masses are outside the range of the \pol grid, and the fitting formulae proposed by \hur do not generalize well for stars outside the range of the grid. There is a second explanation for these discrepancies: at high masses, stars are subject to significant stellar winds (even when they are moderated by a scaling factor of 0.333), and the \hur tracks do not react consistently to mass loss. This point will be discussed in more details in Sect. \ref{mass_loss_impact}.
    \item 
    The maximum radius reached during the evolution differ by one order of magnitude or more. No matter the initial mass, \seba tracks reach the red supergiant (RSG) phase, and extend up to several thousands solar radii. In contrast, the most massive \mesa models stop at some point during their crossing of the Hertzsprung gap and start going back to the blue. As discussed in Sect. \ref{mass_loss_impact}, this is a consequence of the more consistent reaction to mass loss of the \mesa models. Because of this, the maximum extent of the most massive \mesa stars is significantly smaller than that of the \seba models. This has important consequences for the evolution of triple systems, as the occurrence of some interactions (in particular mass transfer) strongly depend on the radial expansion of the stars.
    \item \seba models with $M\ge 70$\,M$_\odot$ end their evolution as WR stars. The corresponding \mesa models do not reach the WR phase and end their evolution as blue supergiants. 
    \item Even with a moderate wind scaling factor, the inconsistent reaction to mass loss of the \hur tracks can be noticed among the more massive stars. A distinct inflection point is observed during the MS evolution of the \seba models. They are marked with a dot on each track. After this inflection point, the luminosity of the stars starts decreasing. The reason of the inflection points and the subsequent decrease in luminosity is a direct consequence of \hur tracks reaction to mass loss. They correspond to the moment the models reach the Vink bistability jump, which makes the mass loss increase. The greater mass loss changes the terminal age main sequence (TAMS) luminosity of the star, which is a simple function of the current mass in the \hur tracks\footnote{Eq. (8) in \huur.}. As a result, the luminosity of the star decreases as its mass decreases, and the track contains a direct imprint of the bistability jump. Although \mesa models' evolution also react to mass loss, they do not show a decrease in luminosity when reaching the bistability jump, which arises in the \seba models when the star shifts towards the track of a star of lower mass.
\end{enumerate}
\subsection{Impacts of mass loss}\label{mass_loss_impact}
\begin{figure}[h]
\centering
\centerline{\includegraphics[trim=0.5cm 0.2cm 1.4cm 1.6cm, clip=true, width=1.0\columnwidth,angle=0]{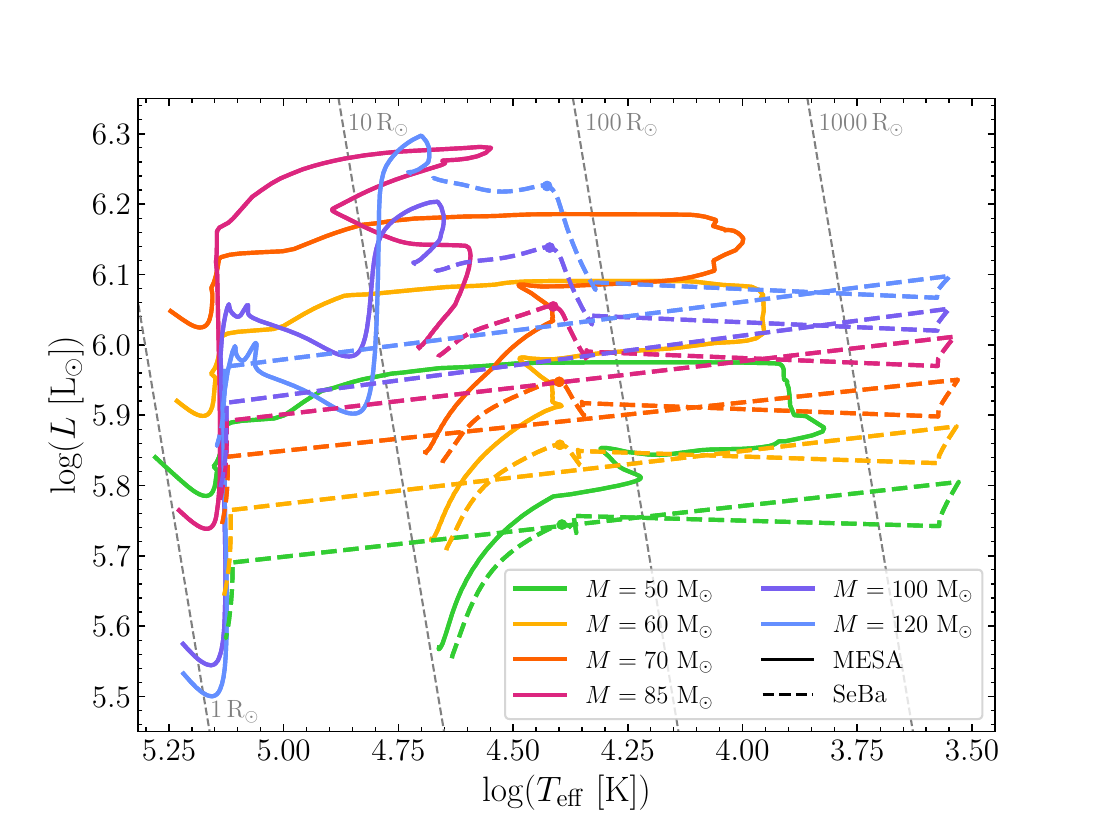}}
\caption{
Same as Fig. \ref{HRD_massive} with a wind scaling factor three times higher.}
\label{HRD_massive_winds}
\end{figure}

In this section, we focus on stars in the upper mass range ($M\in[50,\,120]\,$M$_\odot$) and on the models' respective reactions to mass loss. The HRD for stars in the same mass range as in Fig. \ref{HRD_massive} but with \texttt{Dutch\_scaling\_factor=1} is shown in Fig. \ref{HRD_massive_winds}.
As expected in this high mass range where stellar winds play an important role, increasing the wind scaling factor strongly affects the evolution of the stars. However, the reaction to mass loss is very different between the \mesa and the \seba tracks.

As with the lower wind scaling factor, the maximum radial extent significantly differs between \mesa and \seba tracks, but this difference is even more pronounced. Even with this high mass loss rate, \seba models reach the RSG phase and expand up to several thousands solar radii. In contrast, the maximal radial extent of the \mesa models is strongly reduced in this case, as a result of the significant mass loss. 

With the high wind prescription, \mesa and \seba models all end their evolution as WR in this mass range, but the evolution prior to the WR phase differs. Due to the strength of the winds, the more massive \mesa stars already reach the WR phase during their MS and never expand more than a few tens of solar radii (in the presented models this is the case for the $M=100$\,M$_\odot$ and the $M=120$\,M$_\odot$ stars). This can be explained as follows: when the winds are strong enough, the mass loss timescale becomes smaller than the nuclear timescale. In this case, winds are able to strip the whole hydrogen envelope, which induces the WR phase before the end of the MS. Once the helium enriched layers are revealed, the surface opacity changes which makes the star contract (see \citep{bav23} for a similar discussion). Among the less massive models, the winds are weaker and the stars reach the WR phase later. After the end of their MS, they start crossing the Hertzsprung gap, but come back to the blue and do not become RSGs (for the 85\,M$_\odot$ this happens shortly after the MS and the star's Hertzsprung gap crossing is minor). Their maximal radial extent is lower than with moderate winds.

Unlike \mesa models, \seba models all experience a RSG phase before becoming WR stars. This is because the condition for \seba models to become WR (helium stars) is to lose their whole envelope, i.e. to reach $M_{\rm env}=M_{\rm tot}-M_{\rm core}=0$. This never happens during the MS as the core mass is set to 0 during this stage. At the end of the MS, a core mass is defined, which does not account for the mass already lost during the MS (see Sect. 5.1.2 in \huur). As a result, the envelope mass is overestimated for stars that have undergone significant mass loss due to stellar winds, which delays their transition to the WR phase. The inflection point during the MS evolution of the most massive \seba models, which is caused by the Vink bistability jump and the models' reaction to mass loss is even more pronounced in this case, as the mass loss is larger.

\begin{figure}[h]
\centering
\centerline{\includegraphics[trim=0.5cm 0.9cm 1.2cm 2.1cm, clip=true, width=1.0\columnwidth,angle=0]{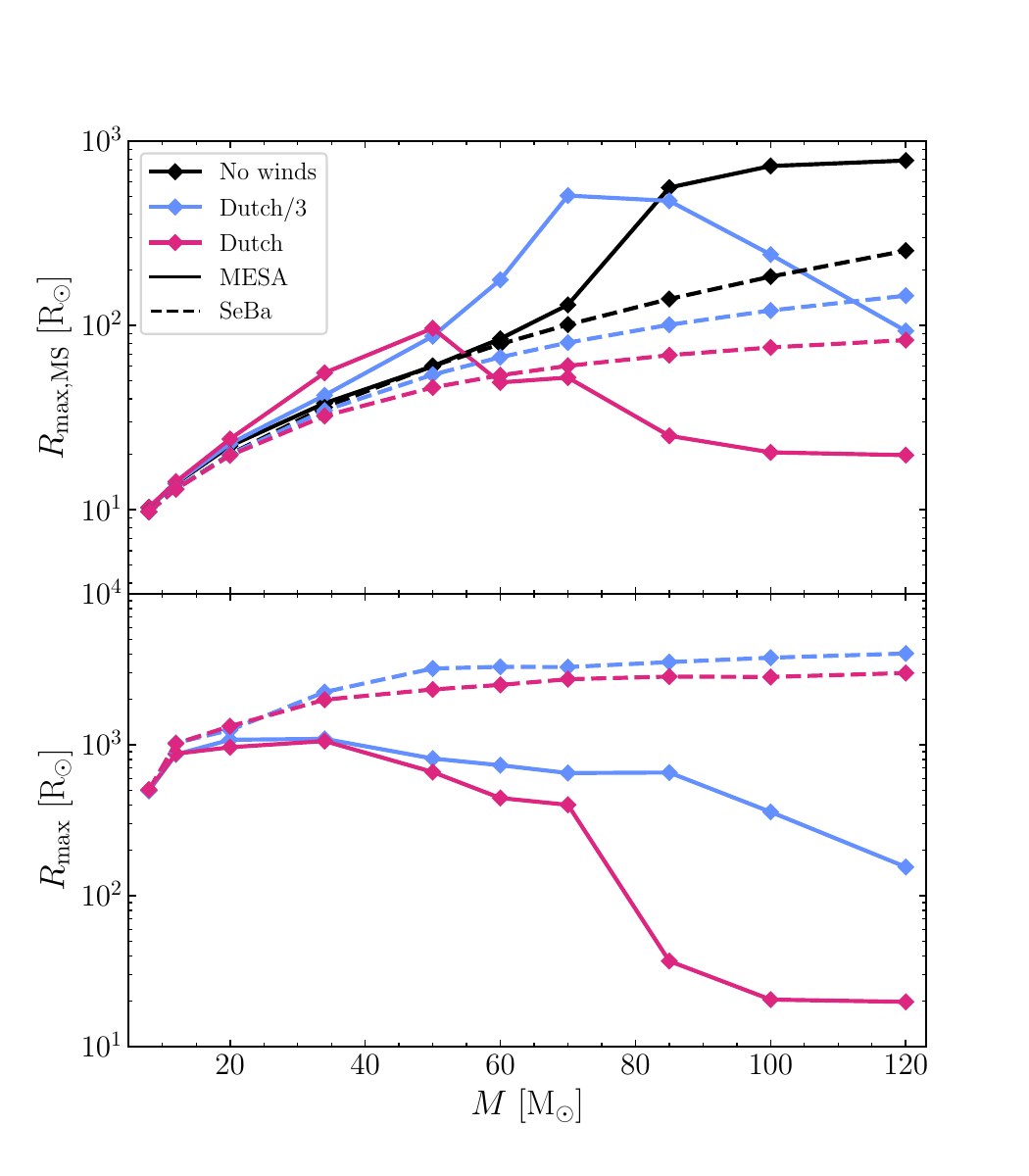}}
\caption{
\textit{Upper panel:} Maximum radius reached during the MS by the \mesa and \seba models for different values of the wind scaling factors (0, 0.333 and 1). \textit{Lower panel:} Same for the maximum radius reached during the whole evolution and for wind scaling factors of 0.333 and 1.}
\label{max_radius}
\end{figure}

The maximum MS radius and the maximum radius play an important role in determining the evolutionary pathways of triple systems, as will be discussed in Sect. \ref{triple_systems}.  The interactions often depend on whether the stellar components fill their Roche Lobe at some point of the evolution, which is mainly determined by the maximum radius they reach.  We therefore show in Fig. \ref{max_radius} the maximum radius reached by \mesa and \seba models during the MS and the whole evolution with different values for the wind scaling factor (for this comparison we also computed MS models without mass loss; we therefore have three values for the scaling factor for the maximum MS radius: 0, 0.333 and 1).

As can be observed in the upper panel of Fig. \ref{max_radius}, the maximum MS radius $R_{\rm max,MS}$ of the \mesa and \seba models show a very different mass and wind scaling factor dependence. The maximum MS radius of the \seba models show a monotonic dependence with the mass, which can be easily understood considering that following the \hur formulae, the TAMS radius is a simple increasing function of the mass\footnote{Eq. (9b) in \huur.}. More interestingly, one notices that for any initial mass, the maximum MS radius of the \seba models decreases when more mass loss is accounted for. This can be simply understood noting that in a \seba run accounting for mass loss, the value of the TAMS radius is updated with the current value of the mass, as is done for the TAMS luminosity. After a star has lost some amount of mass, it shifts towards the track of a lower mass star, and therefore ends up with a smaller TAMS radius, which explains the mass loss dependency of $R_{\rm max,MS}$ for the \seba models.

Detailed calculations with \mesa show that the picture is more complicated. Although models without mass loss show a mass dependence of $R_{\rm max,MS}$ consistent with \seba (i.e. increasing with the mass), the monotonicity of $R_{\rm max,MS}$ with the initial mass breaks down when mass loss is accounted for. With a moderate mass loss (\texttt{Dutch/3}), $R_{\rm max,MS}$ peaks around 70\,M$_\odot$, while with a higher mass loss (\texttt{Dutch}), the peak is around 50\,M$_\odot$. At high masses, the mass loss is so strong that it prevents the stars from expanding.

It is interesting to have a look at the dependence of $R_{\rm max,MS}$ on the wind mass loss for a fixed mass. At 50\,M$_\odot$, the \seba and the \mesa models show an opposite behavior. When no winds are accounted for, the \mesa and \seba models have the same $R_{\rm max,MS}$ (which is due to the of the calibration we performed, see Appendix \ref{AppA}). With greater mass loss, $R_{\rm max,MS}$ of the \mesa model increases, whereas $R_{\rm max,MS}$ of the \seba model decreases. At a higher initial mass (e.g. 70\,M$_\odot$), the mass loss dependence of $R_{\rm max,MS}$ of the \mesa model is non monotonic: the model with moderate winds has a higher maximum MS radius than the model without mass loss, whereas the model with high winds expands less than the model without winds. It is worth noting the convergence of the \seba and the \mesa models for masses smaller than 50\,M$_\odot$, i.e. inside the range of the \pol grid. The excellent match of $R_{\rm max,MS}$ of all \mesa and \seba models with moderate or no mass loss in the range $M\in [8,50]$\,M$_\odot$ validates the performed calibration of $\alpha_{\rm ov}$.

The maximum radius $R_{\rm max}$ reached by the \mesa and \seba models is shown in the lower panel of Fig. \ref{max_radius}. For any model in the considered mass range, the maximum radial expansion predicted by \mesa and \seba substantially differ. The discrepancy between the \mesa and \seba models increases with the mass and the wind scaling factor. This directly results from the inconsistent reaction to mass loss of the \hur tracks. It is more pronounced at high masses as the mass loss is more efficient in this case. As explained above, \seba models experience a RSG phase before becoming WR stars, which explains their large maximum radial extent. \mesa models in contrast react self-consistently to mass loss. Their maximal radial extent is sensitive to the initial mass and to the efficiency of the winds. For the considered range of masses and wind scaling factors, the maximum radial extent strongly decreases with increasing mass and wind scaling factor. In particular, when a wind scaling factor of 1 is applied, we observe a discrepancy of 2 orders of magnitude or more of the maximum radius of \mesa and \seba models with $M\ge 85\,$M$_\odot$. In the mass range $M\in[8,50]\,$M$_\odot$, the \mesa and \seba models show small discrepancies in $R_{\rm max}$, which decrease with the mass. They can be attributed to the winds, which are not completely negligible in this mass range, and increase with the initial mass.
\section{Triple evolution}
\label{triple_systems}
The initial conditions of the different systems presented in this section are given in Table \ref{tab:initial_cond}. The last row indicates whether the evolutionary pathways predicted by \mesa and \seba differ, following the nomenclature adopted in \citep{kum23}\footnote{The possible evolutionary pathways are listed in Sect. \ref{evolutionary_path}.}.
\begin{table*}[h]
\centering
\caption{Initial conditions of the simulated triple systems. }
\begin{tabular}{c c c c c c c c}
\hline \hline
\textbf{Parameters}  & \textbf{$\bm \eta$ Carinae} & \textbf{Tertiary MT} & \textbf{Dyn. Destab.} & \textbf{Orbit unbound} & \textbf{TCO}  & \textbf{ZLK \& precession} & \textbf{$\bm \eta$ Carinae*}\\ 
\hline 
 $M_1$ [M$_\odot$] &110&70& 100  & 85&85&25 & 110\\  
$M_2$ [M$_\odot$] &30&70& 90 & 85 &85&10& 30\\   
$M_3$ [M$_\odot$] &30&80& 25 & 85 &85&30 & 30\\
$a_{\rm in}$ [au] &1&2& 2.5 & 5 &1&6 & 1.1 \\ 
$a_{\rm out}$ [au] &25&25& 25 & 5000 &10&400 & 25 \\ 
$e_{\rm in}$ &0.1&0& 0 & 0 &0&0.1&0.1 \\ 
$e_{\rm out}$ &0.2&0& 0.5 & 0.9 &0&0.3&0.2 \\
$i_{\rm mut}$ [º] &90&0& 0 & 0&0 &90&90\\
$g_{\rm in}$ [rad] &0.1&0& 0 & 0 &0 & 0 & 0.1 \\ 
$g_{\rm out}$ [rad] &0.5&0& 0 & 0 &0& 0 & 0.5 \vspace{2pt} \\  \hline \hline 
\textbf{Divergent} & No & Yes & Yes & Yes & Yes & Yes  & No \\\hline \hline
\end{tabular}
\centering
\tablefoot{The last row indicates whether \mesa and \seba evolutionary pathways differ, following the nomenclature adopted in \citep{kum23}.}
\label{tab:initial_cond}
\end{table*}
\subsection{Validation with $\eta$ Carinae}\label{eta_car_section}
$\eta$ Carinae is an observed binary system with two massive stars ($M_1 \sim 90$\,M$_\odot$ and $M_2 \sim 30$\,M$_\odot$) in a highly eccentric orbit ($e = 0.9$) and period $P = 5.5$\,yr \citep{dam97,dav97}. It is famous for undergoing the "Great Eruption" in the 19th century, during which event the luminosity of the system increased by orders of magnitudes, until it became the second brightest star of the sky \citep{dev56}. The Homunculus Nebula, now surrounding the $\eta$ Carinae system, is an evidence of this tragic event \citep{hum99}. Up to this day, there is no clear consensus on the cause of the Great Eruption. Different scenarios have been proposed, which include mass transfer from $\eta$ Carinae B at periastron passage \citep{kas10}, Luminous Blue Variables (LBV) outbursts \citep[e.g.,][]{hum94}, or the merger of a close inner binary in a hierarchical triple system, leaving behind a binary in a wide orbit \citep{por16,hir21}.

\citet{too16} reproduced the evolution of the favored model of \citep{por16} with \tress. In their simulation, \seba was used for the stellar evolution. The system being composed of very massive components (in their simulation the primary is of initial mass $M_1=110$\,M$_\odot$), the stellar evolution provided by \mesa is expected to differ substantially from that of \seba (see discussion in Sect. \ref{grid_comparison}).

\begin{figure}[h]
\centering
\centerline{\includegraphics[trim=0.3cm 0.7cm 1.5cm 2.2cm, clip=true, width=1.0\columnwidth,angle=0]{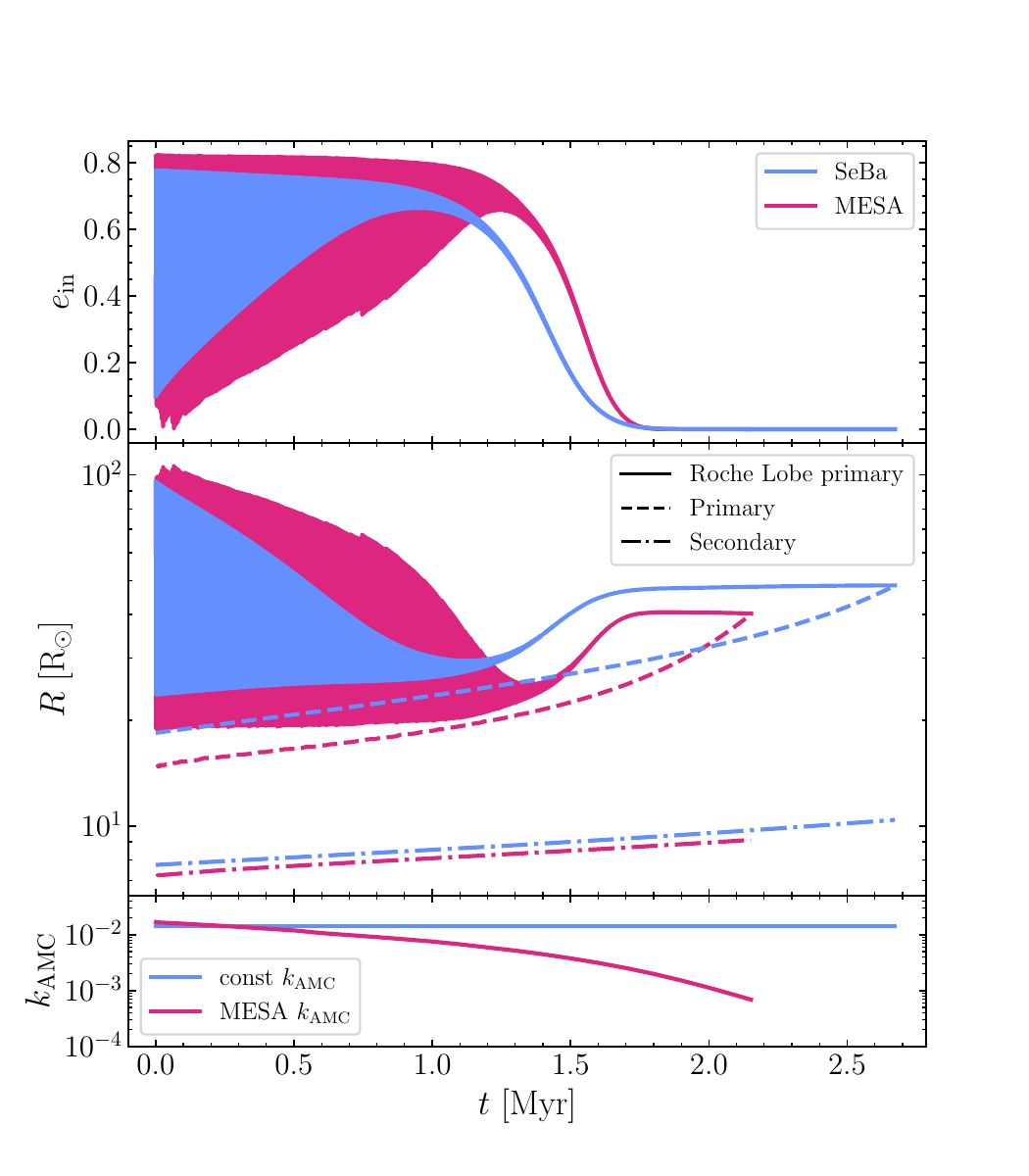}}
\caption{Evolution of the $\eta$ Carinae system simulated with \mesa and \sebaa.
\textit{Upper panel:} Inner eccentricity.  \textit{Middle panel:} Radius expansion of the primary (dashed line) and secondary (dotted \& dashed line). The Roche Lobe radius of the primary is shown in solid line. \textit{Lower panel:} Evolution of the apsidal motion constant of the primary.}
\label{eta_carina}
\end{figure}

We replicated the system using the same initial conditions as in \citep{too16}, i.e. the favored model of \citep{por16}. The system is labeled "$\eta$ Carinae" in Table \ref{tab:initial_cond}.
In Fig. \ref{eta_carina}, the radius expansion of the primary and secondary are shown, together with the Roche Lobe radius of the primary. The evolution of the inner eccentricity and of the apsidal motion constant are also presented. The evolution provided by \seba is compared to that of \mesaa. We find that for this specific system, the evolutionary pathway of the system obtained with \mesa is consistent with \sebaa, i.e. the primary fills its Roche Lobe, which is expected to lead to a common envelope event and eventually a merger of the inner binary, given the high mass ratio of the binary \citep{too16}.
The system begins in a phase of eccentric ZLK oscillations (initially the octupole parameter is $\epsilon_{\rm oct}=0.0048$), which are quenched after $\sim 1.3$\,Myr in the \seba simulation,  $\sim1.5$\,Myr in the \mesa simulation. The quenching of the ZLK oscillations is due to the precession caused by to the distortion and rotation of the stars, which overtakes the precession caused by the ZLK mechanism \citep{too16}. The visible oscillations of the Roche Lobe radius (see Eq. \eqref{roche_lobe}) are a consequence of the ZLK cycles: the Roche Lobe radius is taken at periastron and therefore directly depends on the eccentricity. Subsequently, tides circularize the system (around $t=1.6$\,Myr) and the primary fills its Roche Lobe in a circular orbit. 

The main differences between the outcome of the \mesa and \seba simulations are found in the timescale of the evolution and the radius expansion of the primary. Because the \mesa model expands more during its MS (Sect. \ref{mass_loss_impact}), the star fills its Roche Lobe after only 2.2\,Myr (2.7\,Myr in the \seba run). We further observe that the amplitude of the ZLK oscillations is larger in the \mesa simulation. This difference can be attributed to the fact that in the \mesa run, the apsidal motion constant is computed consistently with the stellar structure, whereas it is assigned a fixed value in the \seba run. As shown in the lower panel of Fig. \ref{eta_carina}, the value at ZAMS of the apsidal motion of the \mesa model is relatively consistent with the value of the \seba model ($k_{\rm AMC}=0.0144$), but at the end of the simulation, it has decreased by more than an order of magnitude. The decrease of $k_{\rm AMC}$ reduces the precession caused by the distortion of the star (Eq. \eqref{prec_tid} and \eqref{prec_rot}), allowing the precession due to the ZLK mechanism to play a more prominent role, which enables the ZLK oscillations to perdure longer. We note that this aspect per se may imply divergences in the evolutionary pathways of triple systems computed with \mesa and \sebaa. In triple systems, the ZLK oscillations typically increase the rate of interactions between the components of the inner binary compared to such interactions in isolated binaries \citep[e.g.,][]{ant17,ham19,too20,ste22,kum23}. By obtaining the apsidal motion constant from the structure, the relative effect of the precession due to ZLK cycles is enhanced, which should increase the number of systems that interact as a result of the ZLK oscillations. In Sect. \ref{div_prec} and in Appendix \ref{AppB}, we present examples of triple systems representative of this aspect.

The differences in the stellar evolution provided by \mesa and \seba are not sufficient to imply a divergent evolutionary pathway for this system, nevertheless it consists of an interesting science case and a proof of concept of our coupling of \mesa to \tress.

\subsection{Divergent evolution: inner or tertiary mass transfer?}
In binary systems, because the primary star is more massive and evolves faster, mass transfer in general occurs from the primary to the secondary \citep[although stellar winds can complicate the picture among very massive stars, see e.g.,][]{bav23}. In the case of triple systems, stable configurations can exist where the tertiary is more massive than each component of the inner binary. One observed system with this property is TIC 470710327 \citep{eis22}. In such systems, mass transfer can theoretically be initiated from the tertiary to the two stars of the inner binary. 

The system presented in this section is labeled "Tertiary MT" in Table \ref{tab:initial_cond}. It is composed of three massive stars, with a tertiary star more massive ($M_3=80$\ms) than the two stars of the inner binary ($M_1=M_2=70$\ms). This system was chosen to represent a class of systems for which a divergent evolutionary pathway is obtained whether \mesa or \seba is used as stellar code and is not unique. We furthermore chose a system not subject to ZLK oscillations ($i_{\rm mut}=e_{\rm out}=0$) to simplify the description.

Given the initial masses of the system, it is not clear whether an inner or tertiary mass transfer will occur. The tertiary being more massive, it evolves faster and has a shorter MS lifetime. It is expected to expand significantly over its MS (and also after the MS in the case of the \seba model, as shown in Fig. \ref{HRD_massive}), which may trigger a mass transfer episode from the tertiary to the inner binary. But since the components of the inner binary are closer, they also possess a smaller Roche Lobe radius, and might therefore interact if the tertiary does not expand enough to fill its Roche Lobe.

\begin{figure}[h]
\centering
\centerline{\includegraphics[trim=0.4cm 0cm 1.8cm 1.7cm, clip=true, width=1.0\columnwidth,angle=0]{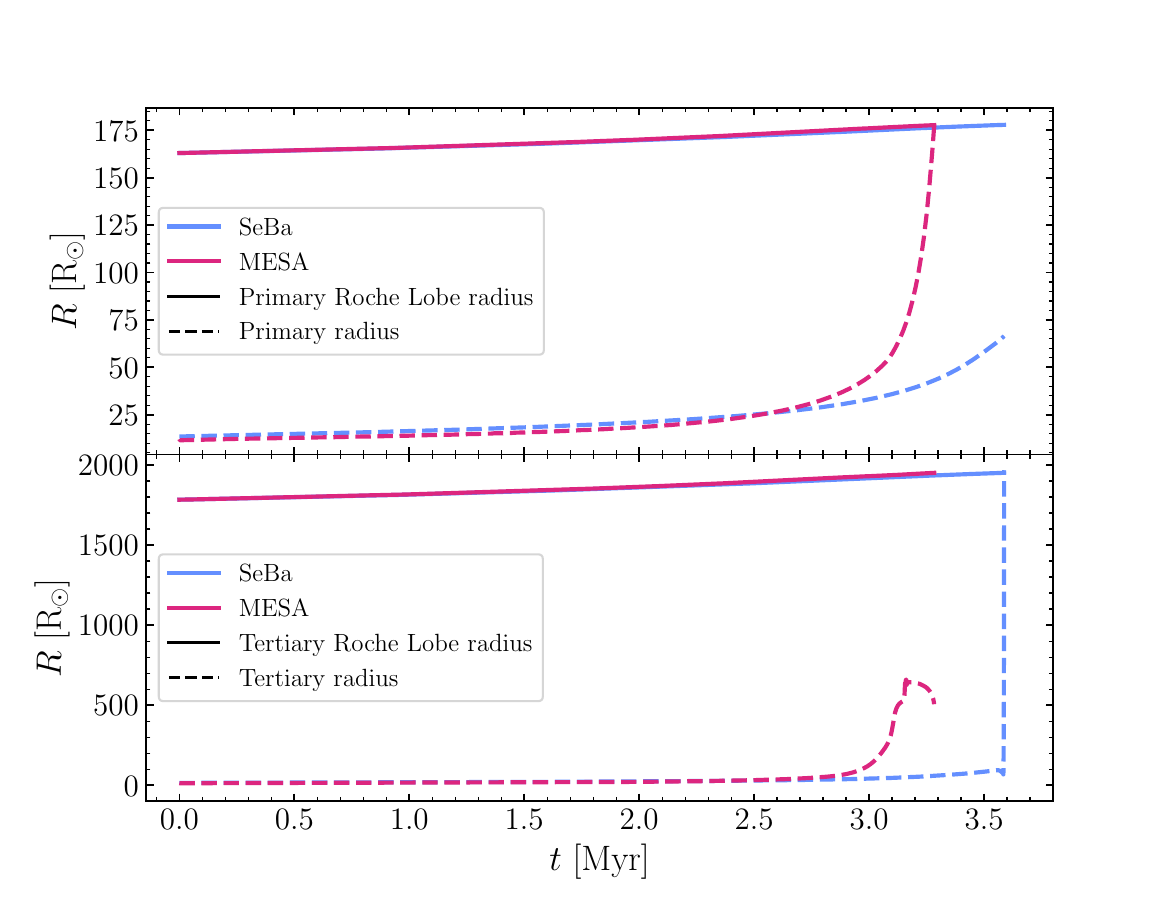}}
\caption{Evolution of the system labeled "Tertiary MT" in Table~\ref{tab:initial_cond}.
\textit{Upper panel:} Radius expansion of the primary (dashed line) and Roche Lobe radius (solid line). \textit{Lower panel:} Same for the tertiary.}
\label{tertiary_mt}
\end{figure}

We simulated this system with \mesa and \seba and compared the obtained evolutionary pathways. The radius of the primary (tertiary) and its Roche Lobe radius evolution are shown in the upper (lower) panel of Fig. \ref{tertiary_mt}. 
In this case, the evolution obtained with the two codes differ significantly. In the \seba simulation, the MS expansion of the components of the inner binary is not sufficient for them to fill their Roche Lobe. As a result, the tertiary completes its MS while the primary and secondary are still far from interacting. It subsequently expands to several thousands solar radii (the maximum radius of the 80\,M$_{\odot}$ \seba model is $R_{\rm max,SeBa}\sim 3500\,$R$_\odot$), becoming a RSG and fills its Roche Lobe ($R_{\rm L}\sim 2000\,$R$_\odot$), which triggers a case B MT to the inner binary. 

The system simulated with \mesa shows a different behavior. After the tertiary completes its MS, it keeps expanding for a while, before starting to contract. It reaches a maximum radius of only $R_{\rm max,MESA}\sim 650\,$R$_\odot$, which is insufficient for it to fill its Roche Lobe. Thus, the two stars of the inner binary end up filling their Roche Lobe, which leads to a contact phase.
\subsection{Divergent evolution: mass transfer or destabilization?}\label{div_destab}
The system presented in this section is labeled  "Dyn. Destab." in Table \ref{tab:initial_cond}. It is composed of three massive stars, with the two components of the inner binary significantly more massive ($M_1=100$\,M$_\odot$, $M_2=90$\,M$_\odot$) than the tertiary ($M_3 = 25$\,M$_\odot$). Initially, it is stable as it has $\frac{a_{\rm out}}{a_{\rm in}}=10$ and $\restriction{\frac{a_{\rm out}}{a_{\rm in}}}{\rm crit}\sim 5.8$. This system represents a class of systems for which divergent evolutionary pathways (inner mass transfer or dynamical destabilization) can be obtained depending whether \mesa or \seba is used as stellar code. 

Being close to dynamical destabilization, the evolution of the system is sensitive to the amount of mass lost through stellar winds. Stellar winds in this mass range being furthermore very uncertain, we simulate the evolution of the system with two winds prescription, the standard \texttt{Dutch} prescription and the \texttt{Dutch} prescription lowered by a factor 3. The use of various winds prescription allows a more robust comparison between the outcome of simulations with \mesa and \sebaa.

\begin{figure}[h]
\centering
\centerline{\includegraphics[trim=0.5cm 0.8cm 1.7cm 2.3cm, clip=true, width=1.0\columnwidth,angle=0]{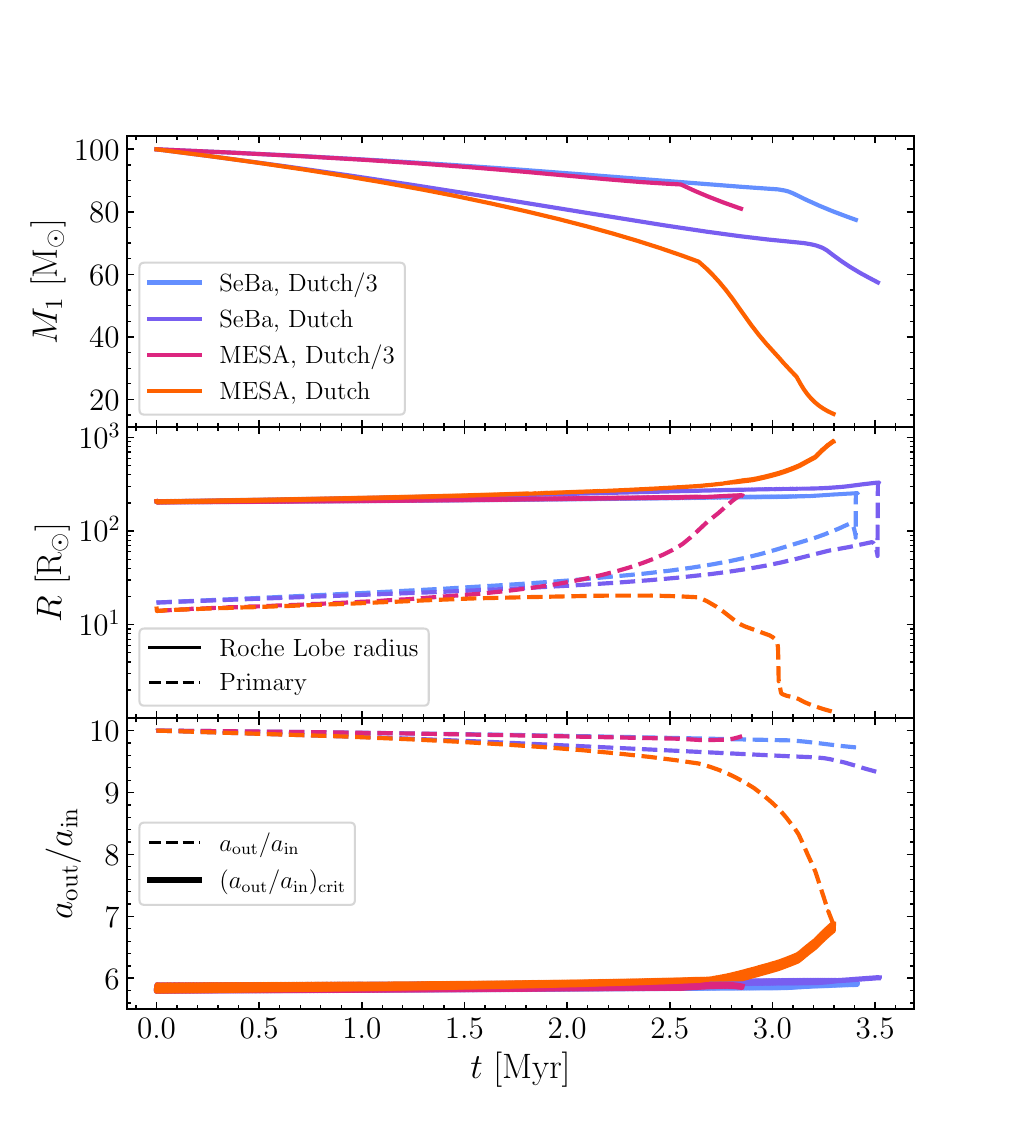}}
\caption{Evolution of the system labeled "Dyn. Destab." in Table~\ref{tab:initial_cond}.
\textit{Upper panel:} Mass evolution of the primary. \textit{Middle panel:} Radius and Roche Lobe radius evolution of the primary. \textit{Lower panel:} Evolution of the ratios $\frac{a_{\rm out}}{a_{\rm in}}$ and $\restriction{\frac{a_{\rm out}}{a_{\rm in}}}{\rm crit}$.}
\label{fig_dyn_destab}
\end{figure}

The evolution of the relevant quantities $\Big($mass, radius, Roche Lobe radius of the primary, ratios $\frac{a_{\rm out}}{a_{\rm in}}$ and $\restriction{\frac{a_{\rm out}}{a_{\rm in}}}{\rm crit}$\Big) of the system obtained with \mesa and \seba with the two winds prescriptions are shown in Fig. \ref{fig_dyn_destab}. 
One can notice that the obtained evolution of the triple system in the \seba simulations are not strongly altered by the wind prescription, as can be expected considering the discussion in Sect. \ref{mass_loss_impact}. In both cases, the primary fills its Roche Lobe after the MS while crossing the Hertzsprung gap (case B MT).

The \mesa simulations are more sensitive to the winds prescription. This can be simply understood looking at the stellar tracks in Sect. \ref{mass_loss_impact}. Indeed, with the standard \texttt{Dutch} prescription, the winds are strong enough for the 100\,M$_\odot$ \mesa star to become a WR already during the MS. Its radius starts shrinking around $t\sim 2.7$\,Myr and never extends beyond $\sim 20$\,R$_\odot$. The star eventually experiences WR winds and the system reaches the dynamical destabilization. With the lowered \texttt{Dutch} winds prescription, the star does not lose enough mass to reach the WR phase during the MS. In this case, the radius expansion is sufficient for the star to fill its Roche Lobe during the MS, which leads to case A MT. 

The system presented in this section is interesting in several aspects. Firstly, it illustrates how the uncertainties in the winds prescriptions can imply uncertainties in the evolution of triple systems. Secondly, it illustrates how an inconsistent reaction to mass loss can impact the predicted evolutionary pathway of a triple system, leading it in one case to a mass transfer, in the other case to a dynamical destabilization. In \citep{bav23}, the same single star physics aspects were discussed, together with the consequences they imply for the evolution of binary systems. They showed that above a certain mass and with the \texttt{Dutch} winds prescriptions, stars computed with \mesa reach the WR phase early in the evolution and as a result do not expand more than a few tens of solar radii. Similar stars computed with \hur tracks based codes (such as \sebaa) become RSGs and expand by several thousands of solar radii. They showed the effect of these important differences on the evolution of binary systems. The natural consequence of this discrepancy is that systems simulated with \hur tracks based codes interact more often at high masses given that the stars expand more.
\subsection{Divergent evolution: mass transfer or unbinding of the orbit?}\label{unbinding}
\begin{figure}[h]
\centering
\centerline{\includegraphics[trim=0.4cm 0.6cm 1.5cm 2.1cm, clip=true, width=0.983\columnwidth,angle=0]{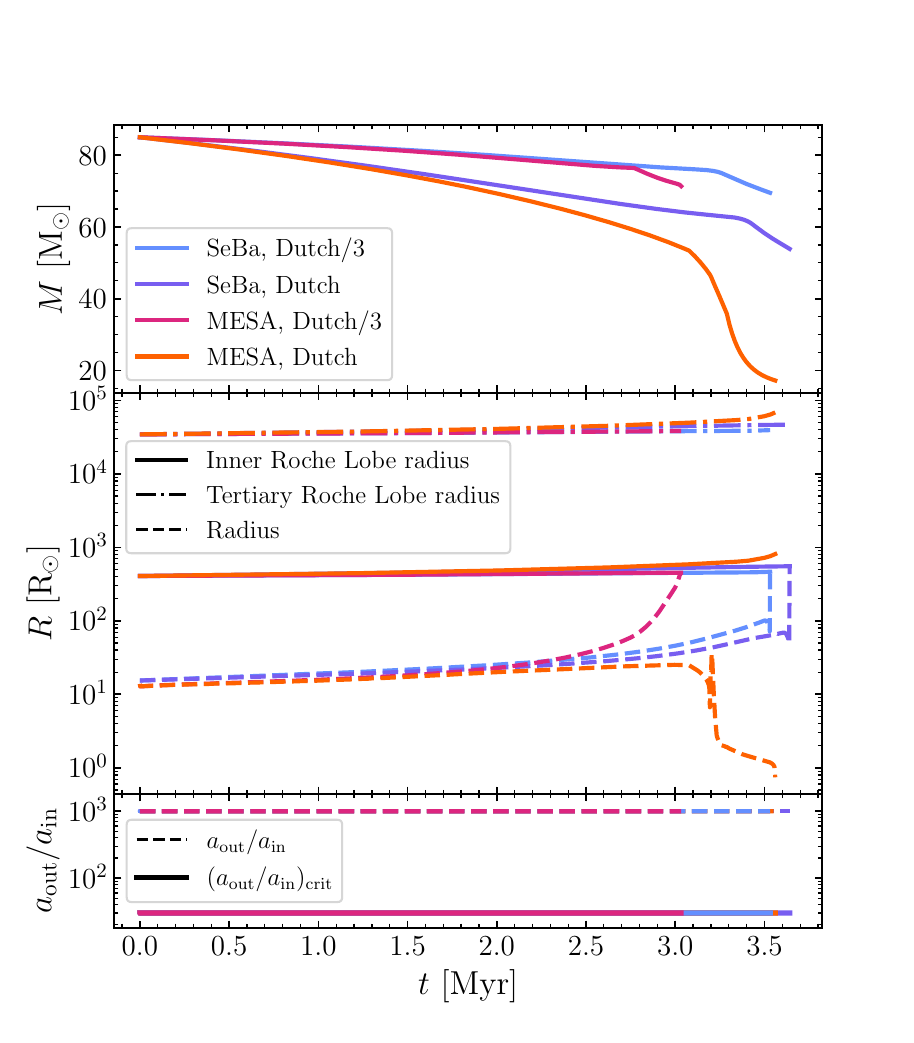}}
\caption{
Same as Fig. \ref{fig_dyn_destab} for the system "Orbit unbound" in Table \ref{tab:initial_cond}.}
\label{orbit_unbound}
\end{figure}

The initial conditions of the system presented in this section are quite close to those of the system of the previous section, except that the chosen masses are smaller and equal ($M_1=M_2=M_3=85$\,M$_\odot$) and the tertiary is  situated at a larger distance. It is labeled "Orbit unbound" in Table \ref{tab:initial_cond}. The initial masses were chosen to be identical for simplicity, but a wide range of initial masses would lead to equivalent results given the largely discrepant values of $R_{\rm max}$ obtained between the two stellar codes (see Fig. \ref{max_radius}). The evolution is shown in Fig. \ref{orbit_unbound}, similarly to Fig. \ref{fig_dyn_destab} with in addition the Roche Lobe radius of the tertiary.
The main difference between this system and that of Sect. \ref{div_destab} is that the present one is initiated far from dynamical destabilization. In this case, stellar winds are not sufficient to destabilize the system. The systems simulated with \seba or with \mesa and moderate winds interact as the primary expands enough to fills its Roche Lobe. In contrast, the components of the system simulated with \mesa and high winds do not expand beyond $\sim 35$\,R$_\odot$. They do not fill their Roche Lobe and thus do not interact before the supernova, similarly to what is found by \citep{bav23} for binary systems. In this case, because the tertiary is on a wide and highly eccentric orbit, it gets unbound from the inner binary as it undergoes the supernova explosion\footnote{Under the assumptions of the supernova dynamics described in the Appendix of \citep{too16}, which is based on \citep{pij12}. It is worth noting that by default, \tres uses the SN kick model of \citep{ver17} in combination with the fallback prescription of \citep{fry12} to scale down the kick velocities for black holes. Whether or not the orbits get unbound depends strongly on the used assumptions for the kick velocity and on the random drawing for its magnitude and orientation, so that repeating the simulation with identical configurations may lead to different outcomes. In general, wide systems are likely to get unbound as they are loosely gravitationally bound.}.
\subsection{Divergent evolution: mass transfer or non interacting system?}
The tertiary of the system presented in Sect. \ref{unbinding} is in a wide outer orbit, which makes it likely to get unbound by the supernova explosion. In the study by \citep{kum23}, only a very small fraction of the simulated systems (0.1-0.4\%) do not experience any of the interactions accounted for in \tres (see Sect. \ref{evolutionary_path}) and end up as triple compact objects systems (TCOs). The reason this category is only scarcely populated is due to a combination of two competing factors. On the one hand, the inner orbit needs to be relatively wide to avoid RLOF. They find that systems with $P_{\rm in}\lesssim  6\times 10^3$\,days cannot avoid mass transfer. This is a direct consequence of the maximal radial expansion of the stellar models (see Fig. \ref{max_radius}). To ensure the stability of the system, this implies that the outer period also needs to be wide, which makes it likely for the tertiary to become unbound due to the SN kick.

Non interacting triple systems are of particular interest in the context of gravitational wave (GW) events as TCO systems are considered to be a main formation channel of mergers \citep[e.g..][]{sil17,ant17,rod18,fra19,mar22,tra22,dor24,su24}. In this scenario, ZLK oscillations caused by the presence of a third object can significantly increase the eccentricity of the inner double compact object (DCO), which can lead it to merge through GW emission.

According to the discussion in Sect. \ref{mass_loss_impact}, the maximum radial expansion of the \mesa models at solar metallicity and with the standard \texttt{Dutch} combination is smaller by one to two orders of magnitudes than that predicted by \seba in the mass range $M\in[50,120]$\,M$_\odot$. Given Kepler 3rd law ($P^2\propto a^3$) and the proportionality between the semi-major axis and the Roche Lobe radius in Eq. \eqref{roche_lobe}, we can estimate the limiting period for avoiding RLOF in \mesa simulations to be up to three orders of magnitude lower than what is predicted by \seba (i.e. $P_{\rm min,MESA}\sim 6$\,days whereas $P_{\rm min,SeBa}\sim 6\times 10^3$\,days), substantially increasing the size of the parameter space where TCOs can form at solar metallicity. Such compact systems are also more likely to stay bound after the SN explosion, as low inner separations allow for low outer separations, i.e. more gravitationally bound systems.

\begin{figure}[h]
\centering
\centerline{\includegraphics[trim=0.5cm 0.cm 1.7cm 1.5cm, clip=true, width=1.0\columnwidth,angle=0]{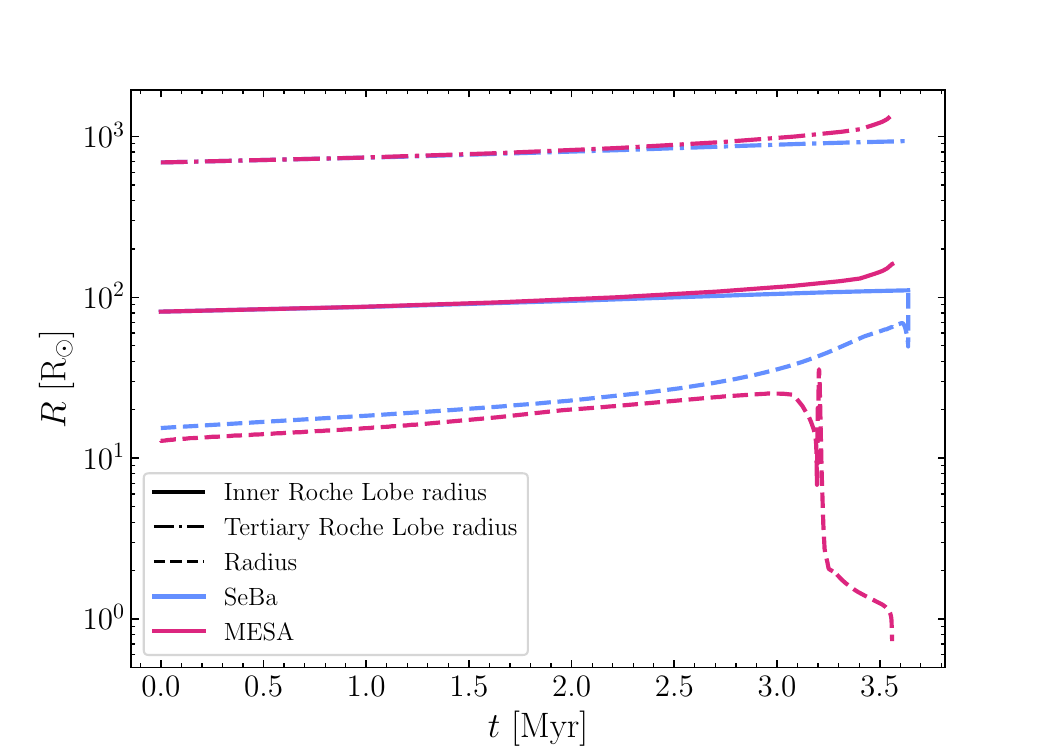}}
\caption{
Same as middle panel of Fig. \ref{orbit_unbound} for the system "TCO" in Table \ref{tab:initial_cond}.}
\label{TCO}
\end{figure}

The system presented in this section is representative of this category and is labeled "TCO" in Table \ref{tab:initial_cond}. It is almost identical to that of Sect. \ref{unbinding}, except that it is much more compact, with $a_{\rm in}=1$\,au and $a_{\rm out}=10$\,au, corresponding to orbital periods $P_{\rm in}=28$\,days and $P_{\rm out}=723$\,days. In this case, the system was only simulated with \texttt{Dutch\_scaling\_factor=1}. The evolution of the radii and Roche Lobe radii of the components are shown in Fig. \ref{TCO}. Since the stars are of the same mass, the radii expansions are the same as in Fig. \ref{orbit_unbound}. Only the Roche Lobe radii differ (they are smaller in this case since the system is more compact). Even in this close inner orbit, the stellar components do not fill their Roche Lobe in the \mesa simulations, as the maximum radius of the $M=85$\,M$_\odot$ model is $R_{\rm max}\sim 35$\,R$_\odot$ and the inner Roche Lobe radius stays larger than 80\,R$_\odot$ throughout the simulation. Given the compactness of the system, it does not get unbound by the SN and forms a TCO system.
The presented system was initiated in a circular orbit and was not subject to ZLK oscillations ($i_{\rm mut}=e_{\rm out}=0)$. It is therefore unlikely to merge by GW emission, but serves as an illustrative example. The formation of GW sources through this channel would be interesting to investigate. Indeed, close systems simulated with \mesa can more easily avoid mass transfer and are also more likely to survive the SN explosion, which should increase the fraction of formed TCOs (and therefore GW progenitors) compared to what is predicted by \citep{ant17,rod18,tra22,kum23}. This aspect will be investigated in a follow-up study.
\subsection{Divergent evolution: the importance of self-consistently modeling the precession}\label{div_prec}
The system presented in this section is labeled "ZLK \& precession" in Table \ref{tab:initial_cond}. It is composed of a tertiary more massive than either star of the inner binary ($M_1=25$\,M$_\odot$, $M_2=10$\,M$_\odot$ and $M_3=30$\,M$_\odot$). The tertiary thus evolves faster, but is situated too far from the inner binary to fill its Roche Lobe. The system is subject to ZLK oscillations. We simulated the system with \mesa and \seba and included a model variation where the stellar evolution is performed by \mesa but the apsidal motion constant is fixed to $k_{\rm AMC}=0.0144$.

The evolution of the relevant quantities (radii of the stars, Roche Lobe radii of the primary and tertiary, apsidal motion constant) are shown in Fig. \ref{precession}.
\begin{figure}[h]
\centering
\centerline{\includegraphics[trim=0.4cm 0.7cm 1.5cm 2.4cm, clip=true, width=1.0\columnwidth,angle=0]{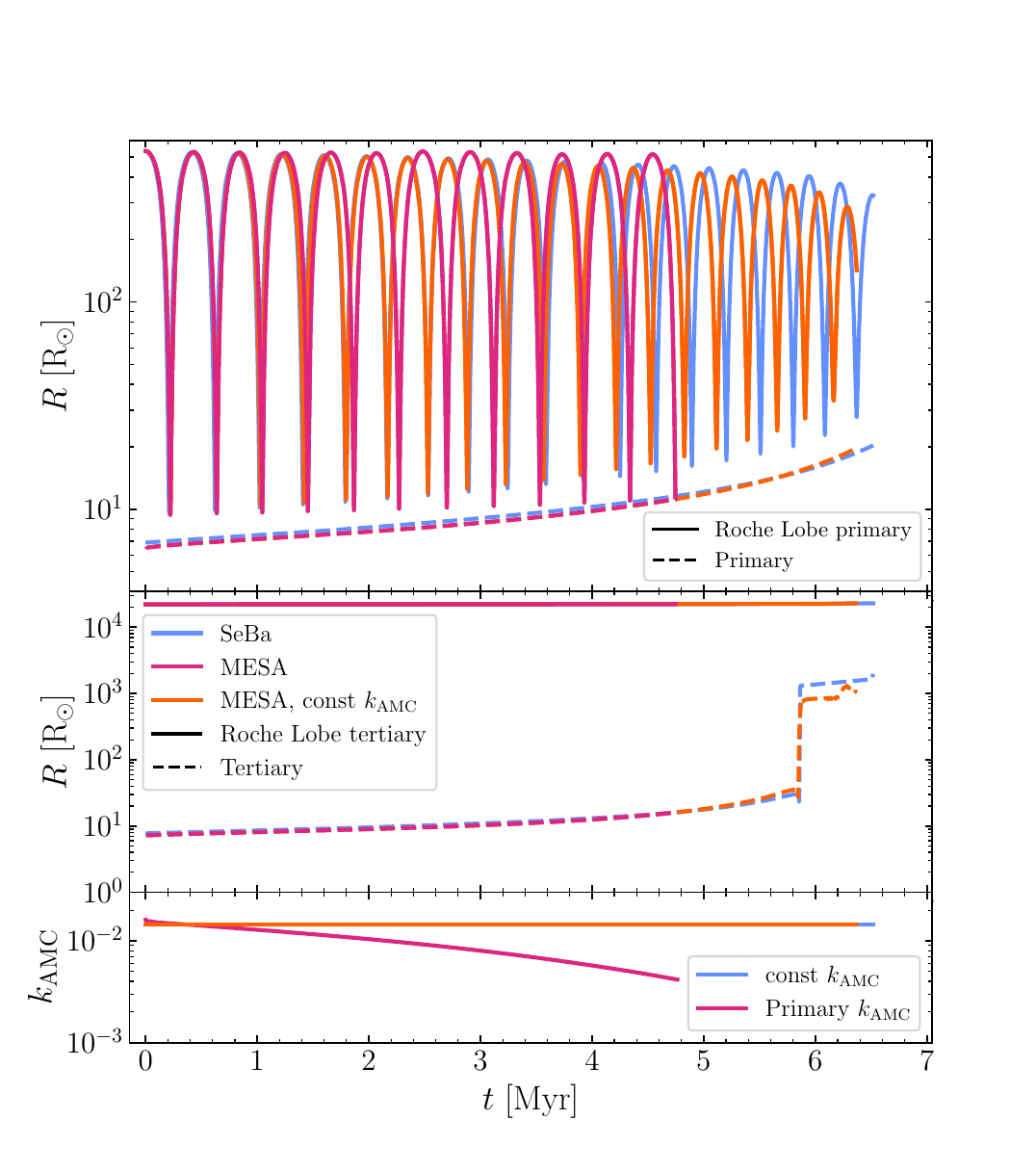}}
\caption{
Evolution of the system labeled "ZLK \& precession" in Table~\ref{tab:initial_cond}.
\textit{Upper panel:} Radius expansion of the primary (dashed line) and Roche Lobe radius (solid line). \textit{Middle panel:} Same for the tertiary. \textit{Lower panel:} Evolution of the apsidal motion constant of the primary.}
\label{precession}
\end{figure}
The system simulated with \mesa and a self-consistent modeling of the apsidal motion constant is subject to maximal ZLK oscillations, as the decrease of $k_{\rm AMC}$ over time makes the precession induced by the ZLK mechanism increasingly dominant. Therefore, the inner eccentricity oscillations stay more pronounced with this model, whereas they are quenched more efficiently with the \seba model and the \mesa model variation.

As a result, the primary fills its Roche Lobe in the case of the \mesa simulation, which leads to case A eccentric mass transfer. In the \mesa model variation and the \seba simulation, the quenching of the oscillations prevents the primary from filling its Roche Lobe. The tertiary eventually completes its evolution, undergoes a SN explosion and gets unbound from the inner binary. It is worth noting that in this case, the chosen initial masses lie inside the range of the \pol grid, and the stellar evolution predicted by \seba is very consistent with that of \mesaa. The divergence in the evolutionary pathway can thus be solely attributed to the apsidal motion modeling.
\subsection{ZLK cycles and tidal evolution}\label{ZLKTF}
In this section we aim to illustrate the impact of the dynamical tides prescription proposed in \sci (adapted from \zahh) on the evolution of triple systems and show how it differs from that of \huu. For this purpose we present four triple systems where the inner binary is in an eccentric orbit close enough to experience substantial tides. Two systems reach an inner eccentricity $e_{\rm in,max}\gtrsim 0.8$ during the evolution, either due to ZLK oscillations or because it is initially high, whereas the inner eccentricity of the other two verifies $e_{\rm in,max}\lesssim 0.6$ throughout the whole evolution. In each category, one system is subject to ZLK oscillations, the other is not. The systems are labeled according to the maximum inner eccentricity they reach and whether they experience ZLK oscillations. The initial conditions of these systems are given in Table \ref{tab:initial_cond_tides}.
\begin{table*}[h]
\centering
\caption{Initial conditions of systems with an eccentric inner binary subject to dynamical tides.}
\begin{tabular}{c c c c c}
\hline \hline
\textbf{Parameters}  & \textbf{No ZLK, \bm $e_{\rm in,max}\lesssim 0.6$} & \textbf{ZLK, \bm $e_{\rm in,max}\lesssim 0.6$} & \textbf{No ZLK, \bm $e_{\rm in,max}\gtrsim 0.8$} & \textbf{ZLK, \bm $e_{\rm in,max}\gtrsim 0.8$}  \\ 
\hline 
 $M_1$ [M$_\odot$] &100&100& 100  & 100\\  
$M_2$ [M$_\odot$] &10&10& 10 & 10 \\   
$M_3$ [M$_\odot$] &10&10& 10 & 10 \\
$a_{\rm in}$ [au] &1&1& 1 & 1 \\ 
$a_{\rm out}$ [au] &20&20& 20 & 20 \\ 
$e_{\rm in}$ &0.3&0.1& 0.8 & 0.1 \\ 
$e_{\rm out}$ &0&0.1& 0 & 0.1 \\
$i_{\rm mut}$ [º] &0&60& 0 & 90 \\
$g_{\rm in}$ [rad] &0&0& 0 & 0 \\ 
$g_{\rm out}$ [rad] &0&0& 0 & 0  \\
\hline \hline
\end{tabular}
\centering
\label{tab:initial_cond_tides}
\end{table*}
All systems are simulated with \sebaa. For each set of initial conditions, two independent simulations are performed, one time with the default \hu prescription, the other with the \sci prescription. A detailed description of the latter model is given in Appendix \ref{AppD} for the interested reader. The main conclusion is that in highly eccentric orbits ($e\gtrsim 0.7$), this tidal model predicts the eccentricity of the system to increase, which goes against the common picture that tides always act towards the circularization of the system\footnote{This result directly stems from \zah formalism of dynamical tides, see Appendix \ref{AppD} for more details.}.

\begin{figure}[h]
\centering
\centerline{\includegraphics[trim=.3cm 0.4cm 1.4cm 1.4cm, clip=true, width=0.98\columnwidth,angle=0]{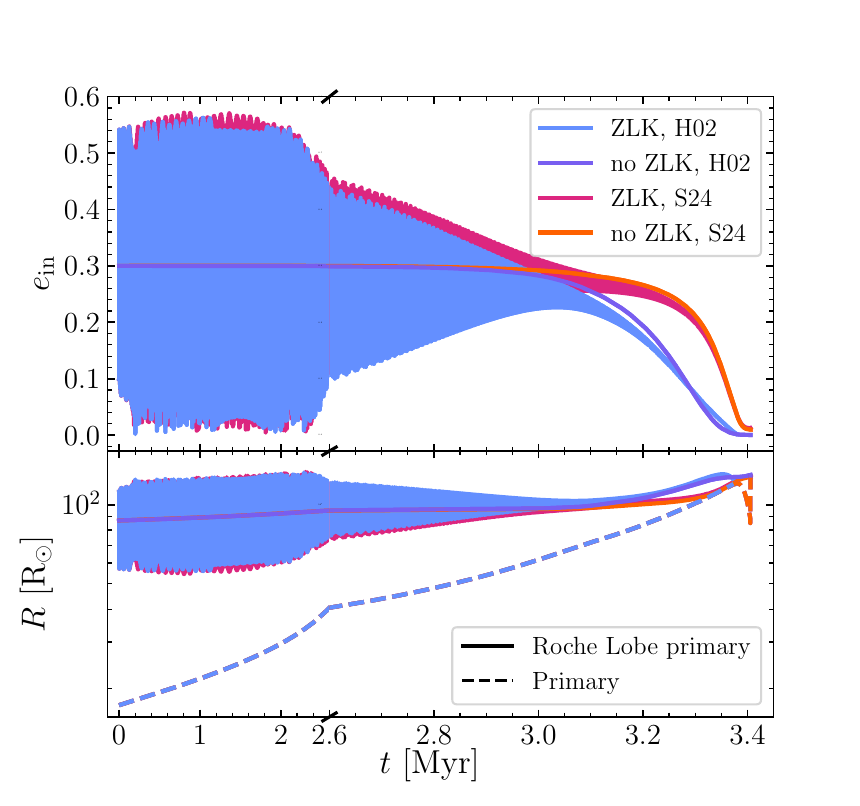}}
\caption{
Comparison of the evolution of systems in moderately eccentric orbits as predicted by the \hu and the \sci prescriptions. One system is subject to ZLK oscillations (blue and magenta lines), the other is not (violet and orange lines). The scale of the horizontal axis changes at $t=2.6$\,Myr and is chosen as to focus on the part of the evolution where tidal effects are important. \textit{Upper panel}: inner eccentricity evolution. \textit{Lower panel}: radius and Roche Lobe radius evolution of the primary.}
\label{tides_low}
\end{figure}

The evolution of the two systems whose inner eccentricity remains lower than 0.6 throughout the evolution is shown in Fig. \ref{tides_low}. They correspond to the two first systems in Table \ref{tab:initial_cond_tides}. One of them is subject to ZLK oscillations, the other is not. As most of the tidal evolution occurs near the end of the simulation, the largest part of the plot focuses on the last 0.8\,Myr of evolution.
For these systems, the evolutions predicted by the two prescriptions are relatively similar\footnote{The evolution of the considered quantities ($e_{\rm in}$, $a_{\rm in}$, $\Omega_1$ and $\Omega_2$) differ according to the chosen prescription, but they overall show the same trend.}. Both predict that the tides circularize the system before the primary fills its Roche Lobe, which implies that the system experiences mass transfer in a circular orbit. This is because in this "low eccentricity regime", the \sci prescription also predicts that dynamical tides tend to reduce the eccentricity (for any pseudo-synchronized star).

\begin{figure}[h]
\centering
\centerline{\includegraphics[trim=0.2cm 0.8cm 1.7cm 2.3cm, clip=true, width=1\columnwidth,angle=0]{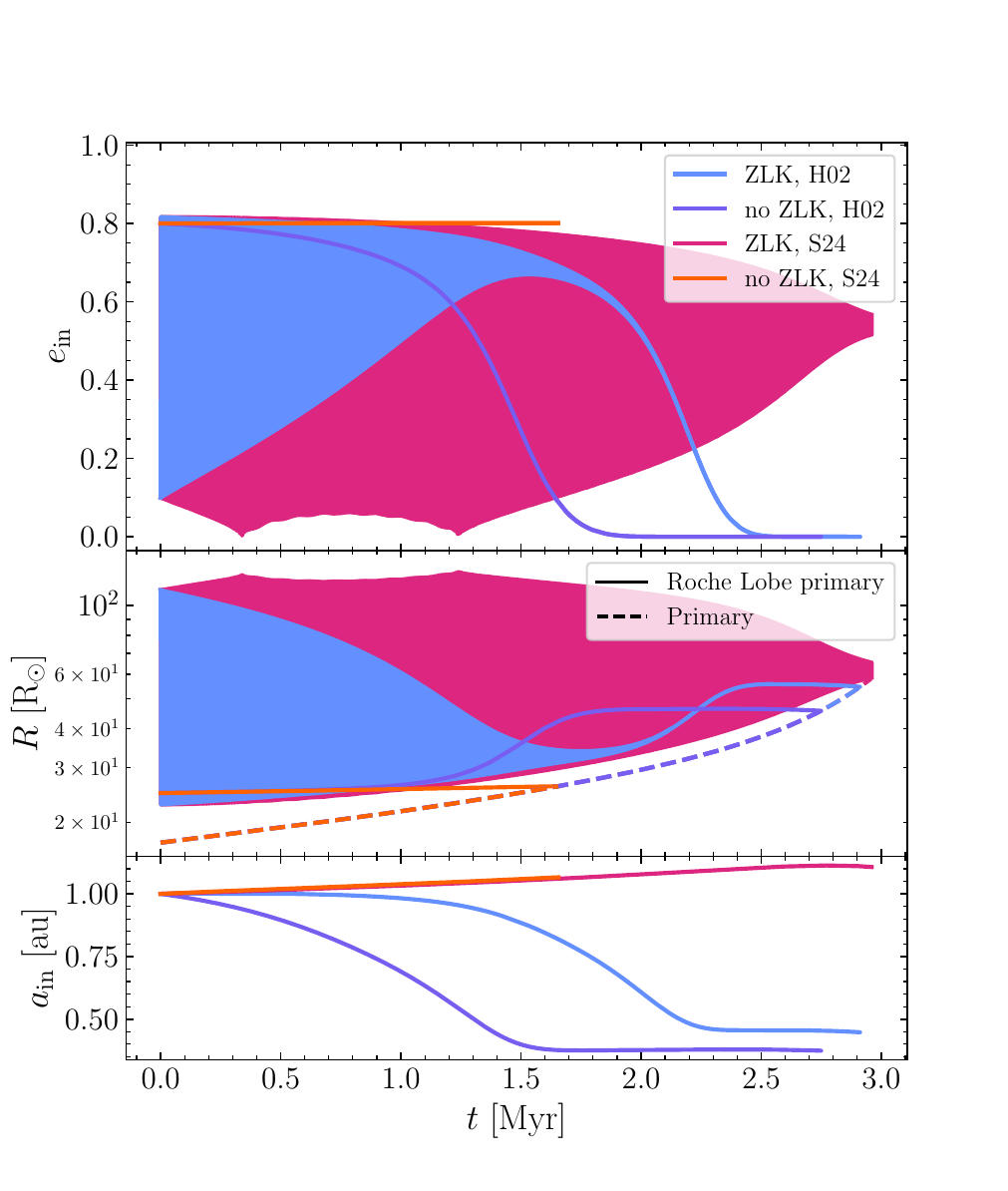}}
\caption{
Same as Fig. \ref{tides_low} for systems with $e_{\rm in,max}\gtrsim 0.8$. Additionally the evolution of the inner separation is shown in the lower panel.}
\label{tides_high}
\end{figure}

The evolution of the two systems whose inner eccentricity reach 0.8 or more during the evolution is shown in Fig. \ref{tides_high}. They correspond to the last two systems in Table \ref{tab:initial_cond_tides}.
For these systems, the tidal evolution predicted by the \sci prescription differs substantially from that predicted by the \hu prescription. In this regime, \sci predicts that the dynamical tides make the eccentricity increase\footnote{In the presented simulations the increase of $e_{\rm in}$ due to tidal interactions is only marginal as it goes along with the increase of $a_{\rm in}$, which quickly quenches the effect of the tides because of the high $R/a_{\rm in}$ dependencies (negative feedback loop). Systems simulated with the \hu prescription show a much more pronounced eccentricity evolution as the decrease of $e_{\rm in}$ goes along with the decrease of $a_{\rm in}$, which strongly increases the strength of the tides (positive feedback loop).}. The systems in this case do not circularize, therefore they experience an eccentric mass transfer episode as the primary fills its Roche Lobe. In contrast, the systems simulated with the \hu prescription circularize before the primary fills its Roche Lobe and experience mass transfer in a circular orbit.

The question of whether the tidal evolution of largely eccentric systems as predicted by \sci (i.e. where the dynamical tides make the eccentricity increase) is justified physically is beyond the scope of this study. We recall that this result directly stems from \zah formalism, which is an expansion keeping only low order eccentricity terms. It is possible that pushing the expansion to higher order terms may alter its prediction, in particular the shape of the function $f_e$, which is crucial for determining the eccentricity evolution (see Appendix \ref{AppD}). When combined with the physics of triple systems (in particular ZLK oscillations), the use of the \sci prescription obviously tends to increase of the number of interacting systems, as any system reaching an inner eccentricity $e_{\rm in}\gtrsim 0.7$ through ZLK oscillations and subject to dynamical tides would not circularize and be therefore more likely to interact, according to Eq. \eqref{roche_lobe}.
\section{Conclusion}\label{conclusion}
In this study we presented the first simulations of triple systems performed with a detailed, on-the-fly stellar code, \mesa and a triple system secular evolution code, \tress. The coupling of \tres to \mesa was done through \amusee. This coupling was relatively straightforward due to the fact that \tres is natively developed within \amuse and thanks to the versatility of this environment. These simulations offer new insights into the physics of triple systems.

The approach we adopted was to primarily investigate the impact of the single star physics on the evolution of triple systems. We kept the same physics for the interactions and investigated how the differences between the stellar evolution predicted by a detailed code such as \mesa and that of a fast stellar code based on the \hur fitting formulae (\sebaa) can lead to divergences in the predicted evolutionary pathways of triple systems.

We primarily focused on stars with masses $M\ge50\,$M$_\odot$, i.e. outside the range of the \pol grid, on which \hur tracks are based. Above this limit the fitting formulae become increasingly less accurate with increasing masses as they extrapolate the tracks. We compared single star evolutionary tracks obtained with \mesa to those obtained with \seba in the mass range $M\in[8,120]$\,M$_\odot$. We showed that \seba and \mesa tracks deviate significantly in the upper end of our mass range. In particular, the maximum radii reached during the evolution, which are crucial for determining whether components of multiple systems interact, already differ by an order of magnitude or more among the most massive stars with a moderate mass loss prescription .

Given the importance and uncertainties of stellar winds, especially for the most massive stars, we investigated the impact of the strength of the stellar winds on our results. We showed that the tracks predicted by \mesa and \seba become increasingly divergent with increasing mass loss, which is a direct consequence of the fact that the \hur tracks do not offer a self-consistent reaction to mass loss. In particular, with a standard winds combination \citep[i.e.][]{dej88,nug00,vin01}, the predicted maximum radii reached during the evolution differ by up to two orders of magnitude depending on the used stellar code in the mass range we considered.

We also provided illustrative examples of triple systems for which the stellar evolution discrepancies between the two codes imply divergences in the evolutionary pathway of the system. We were able to identify examples of divergent pathways for each of the possible evolutionary pathways considered in \tress. These results illustrate the importance of a consistent modeling of the stellar evolution for predicting the evolutionary pathways of triple systems.

We showed that these discrepancies can be primarily attributed to the different radial expansion predicted by the codes, which is a consequence of the different reaction to mass loss they offer and the fact that at masses larger than 50\,M$_\odot$, \hur fitting formulae extrapolate the tracks of lower mass stars. Since \mesa models expand less at high masses, the size of the parameter space where RLOF can be avoided is increased. At solar metallicity, the limiting period to avoid RLOF with \seba models is $P_{\rm min,SeBa}\sim 6\times 10^3$\,days whereas with \mesa it is reduced to $P_{\rm min,MESA}\sim 6$\,days. This has important consequences for the predicted evolutionary pathways of triple systems. In particular, we demonstrated that the parameter space where triple compact objects can form is increased, which can have significant consequences in the context of the progenitors of GW events. The determination of the BH merger rate requires carrying out a population synthesis study, which was beyond the scope of this project, but could be investigated in a follow-up study.

Using a detailed stellar evolution code also allows to self-consistently model some important interactions between the stellar components, in particular the precession of the orbits due to the tidal and rotational distortion of the stars. This precession plays an important role for the evolution of triple systems, as it competes with the precession driven by the ZLK oscillations. In our \mesa simulations, we directly obtained the apsidal motion constant from the structure, which allowed us to self-consistently model the precession due to the distortion of the star and its competition with that driven by the ZLK oscillations. In \sebaa, in contrast, the apsidal motion constant remains fixed and is updated only when the star transitions to a new evolutionary stage\footnote{In the mass range considered in this work (see Sect. \ref{tresmesa})}. We showed that MS models with a self-consistent precession modeling tend to experience more pronounced ZLK oscillations, making them more likely to interact. This is because the apsidal motion constant decreases by orders of magnitude during the MS evolution, reducing the effect of precession caused by stellar distortion and thereby increasing the influence of ZLK oscillations.

Finally, we illustrated the impact of using another prescription for the dynamical tides, as proposed by \scii, on the evolution of a small set of triple systems, and discussed the implication of this prescription when applied to largely eccentric systems.

In summary, using a more self-consistent modeling of the single star evolution alters the predicted evolutionary pathways of triple systems. The great advantage of stellar codes such as \seba is their computing efficiency, which makes them an essential tool for population synthesis studies. The present study complements that of \citep{kum23}, who provided predictions of the main evolutionary pathways of massive triple systems through a population synthesis approach. Quantifying the effect of these discrepancies at a population level requires carrying out a population synthesis study, which would be computationally expensive if \mesa were to be used given the large dimensionality of the parameter space of triple systems, but could be the subject of a follow-up study.
\begin{acknowledgements}
We sincerely thank the anonymous referee for their constructive feedback and valuable suggestions, which enabled us to improve the manuscript. LS thanks Prof. Georges Meynet (UNIGE), Dr. Max Briel (UNIGE) and Dr. Patrick Eggenberger (UNIGE) for useful discussions, and Dr. Sophie Rosu (UNIGE) for a valuable feedback. LS and SE have received support from the SNF project No 212143. LS has received support from the Netherlands Research School for Astronomy (NOVA). SE and SR have received funding from the European Research Council (ERC) under the European Union's Horizon 2020 research and innovation program (grand agreement No 833925, project STAREX). EF is support by SNF grant number 200020\_212124. ST acknowledges support from the Netherlands Research Council NWO (VIDI 203.061 grants). 
\end{acknowledgements}


\bibliographystyle{aa}
\bibliography{myrefs}\begin{appendix}
\section{Calibration of $\alpha_{\rm ov}$}\label{AppA}
\begin{figure}[h]
\centering
\centerline{\includegraphics[trim=0cm 0cm 1.2cm 1.2cm, clip=true, width=1.02\columnwidth,angle=0]{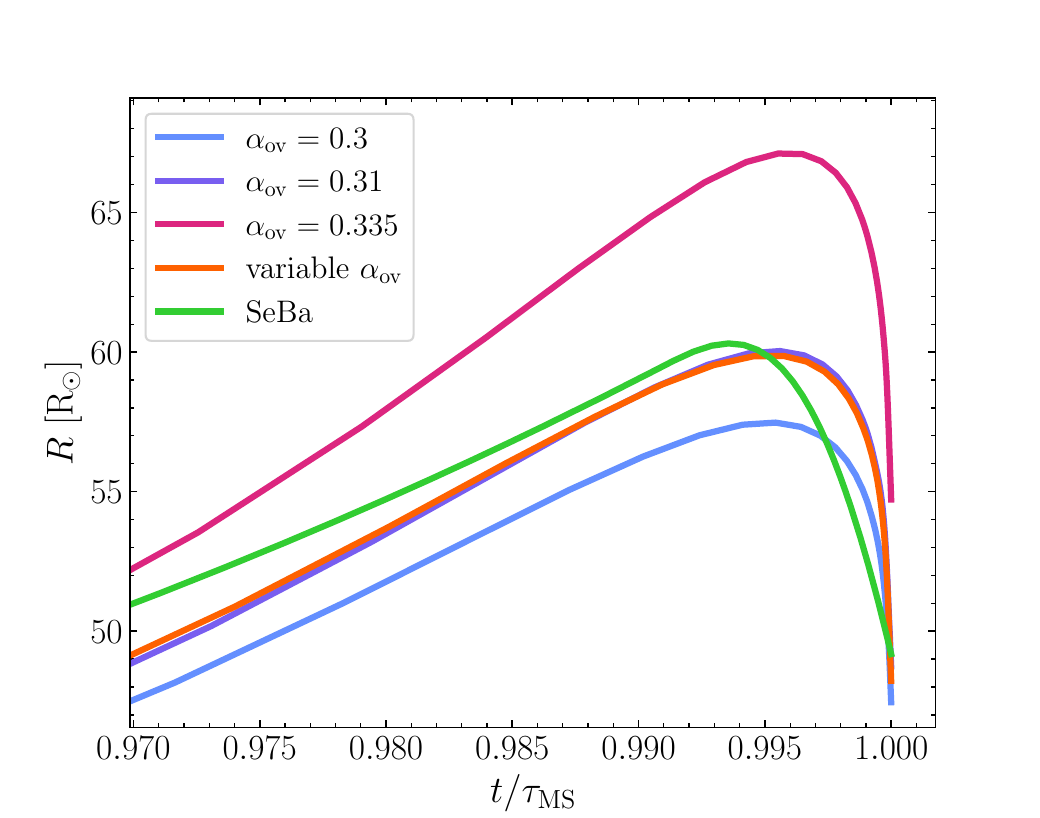}}
\caption{
End of MS radial expansion of the 50\,M$_{\odot}$ \mesa model with different step-overshoot prescriptions compared with the \seba model (all models are without stellar winds). The value $\alpha_{\rm ov}=0.31$ offers the best fit to the \seba model radial expansion. All simulations were computed without mass loss. A toy model with decreasing values of $\alpha_{\rm ov}$ over time as to mimic the effect of the $\delta$ prescription was also computed (orange curve).}
\label{alpha_ov_calib}
\end{figure}

In order to perform our $\alpha_{\rm ov}$ calibration, we computed several \mesa models of 50\,M$_{\odot}$ stars with various $\alpha_{\rm ov}$ values and compared them to the \seba model (all models were performed without accounting for mass loss by stellar winds). As an illustration, we show in Fig. \ref{alpha_ov_calib} the end MS evolution of \mesa models with $\alpha_{\rm ov}=\{0.3,0.31,0.335\}$ (the latter value corresponds to that used in \citep{bro11} which was calibrated using data of the FLAMES survey of massive stars) and compare them to the \seba model.
We find  that the value $\alpha_{\rm ov}=0.31$ offers the best fit to the \seba model in terms of the maximal MS radial expansion. We also computed a model variation were we manually reduced the value of $\alpha_{\rm ov}$ along the MS as to mimic the effect of the $\delta$ prescription used in \pool. With this model, we started the MS evolution with $\alpha_{\rm ov}=0.4$, which corresponds to the ZAMS value stemming from the $\delta$ prescription for this mass (see Fig. 1 in \pool) and subsequently reduced it. More precisely, we set the overshooting parameter value to $\alpha_{\rm ov}=\{0.4,0.35,0.3,0.25\}$ when the central hydrogen mass fraction reached $X_{\rm c}=\left\{X_{0},\frac{3}{4}X_{0},\frac{1}{2}X_{0},\frac{1}{4}X_{0}\right\}$ with $X_{0}=0.7$ the initial hydrogen mass fraction. The MS radial evolution of this model is almost undistinguishable from that of the model with $\alpha_{\rm ov}=0.31$.

\begin{figure}[h]
\centering
\centerline{\includegraphics[trim=.6cm 0cm 1.2cm 1.2cm, clip=true, width=1.02\columnwidth,angle=0]{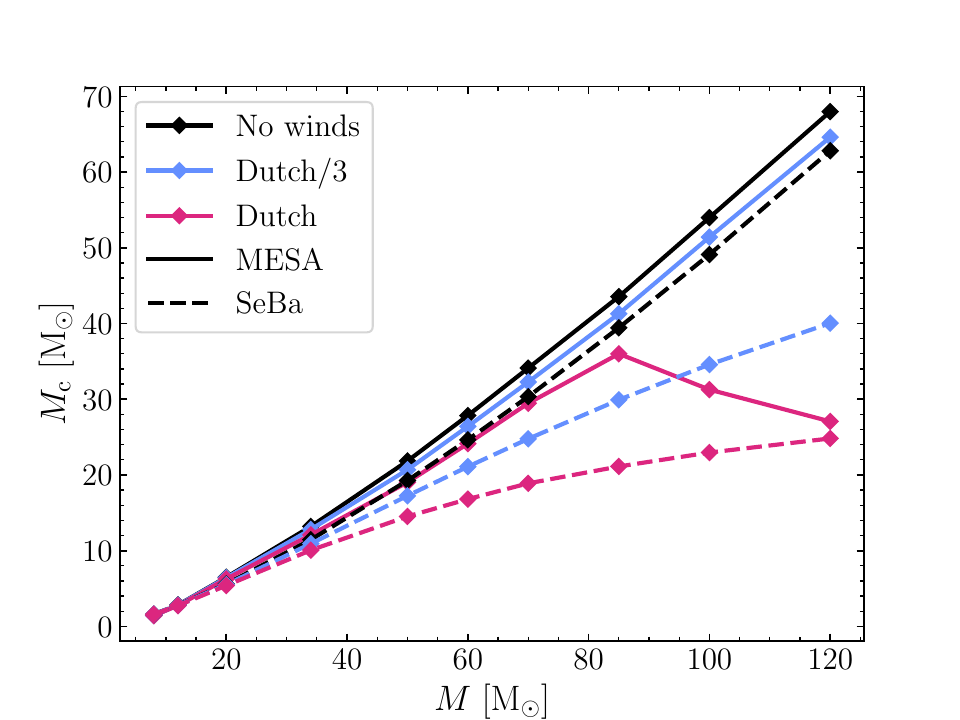}}
\caption{
He core masses at He burning ignition of \mesa and \seba models as a function of mass and with the three different wind scaling factors (0, 0.333 and 1).}
\label{core_mass}
\end{figure}

The mass of the helium core formed during the MS evolution of the \mesa models directly depends on the performed calibration of $\alpha_{\rm ov}$. It is interesting to compare the helium core mass resulting from this calibration to that of the \seba models. Following the \hur fitting formulae, the core mass at helium ignition is given by their Eq. (44), replacing $M_{\rm c}\left[L_{\rm BGB}(M_{\rm HeF})\right]$ with $M_{\rm c}\left[L_{\rm HeI}(M_{\rm HeF})\right]$. In the mass range considered in this work, this formula is very well approximated by 
\begin{equation}
M_{\rm c,HeI}=0.098M^{1.35},
\label{core_mass_seba}
\end{equation}
where $M$ is the current mass of the star in solar mass (see comments below their Eq. (44) and (65)). We show in Fig. \ref{core_mass} a comparison of the He core masses at He burning ignition of \mesa and \seba models as a function of initial mass and with the three different wind scaling factors (0, 0.333 and 1).
We note that when no winds are accounted for, the \seba and \mesa core masses show a consistent trend with the mass, which is expected. An offset between the core mass of the \mesa and \seba models can be noticed, which increases with the mass. The core masses only match for models with  $M\le 20$\,M$_\odot$. Notably, the \mesa and \seba core masses of the 50\,M$_\odot$ model, which is the one we used for calibrating $\alpha_{\rm ov}$, differ by about 10\%. This discrepancy can primarily be attributed to the different treatment of convection between the two approaches. Secondly, the calibration was performed matching the maximum radius, and not the luminosity, which is more directly linked to the mass of the core. One can note looking at the HRD (Fig. \ref{grid_comp_both}) that the \mesa models in general have a slightly higher luminosity when crossing the Hertzsprung gap, which is consistent with them having slightly larger cores. 

Although the core masses are relatively consistent when no winds are accounted for, the discrepancy increases with increasing mass loss. The reason is that the mass used in Eq. \ref{core_mass_seba} is the current mass, not the initial mass. At high masses, the stars have already lost a significant portion of their mass at helium ignition, which is why Eq. \eqref{core_mass_seba} predicts much smaller core masses. Finally, the drop of $M_{\rm c}$ of the 100 and 120\,M$_\odot$ \mesa models is explained by the fact that they loose all of their hydrogen envelope before the end of the MS. At this point, they start losing mass in regions enriched in helium, which reduces the size of their final helium core.
\section{$\eta$ Carinae*}\label{AppB}
\begin{figure}[h]
\centering
\centerline{\includegraphics[trim=0.5cm 0.5cm 1.7cm 2.4cm, clip=true, width=1.02\columnwidth,angle=0]{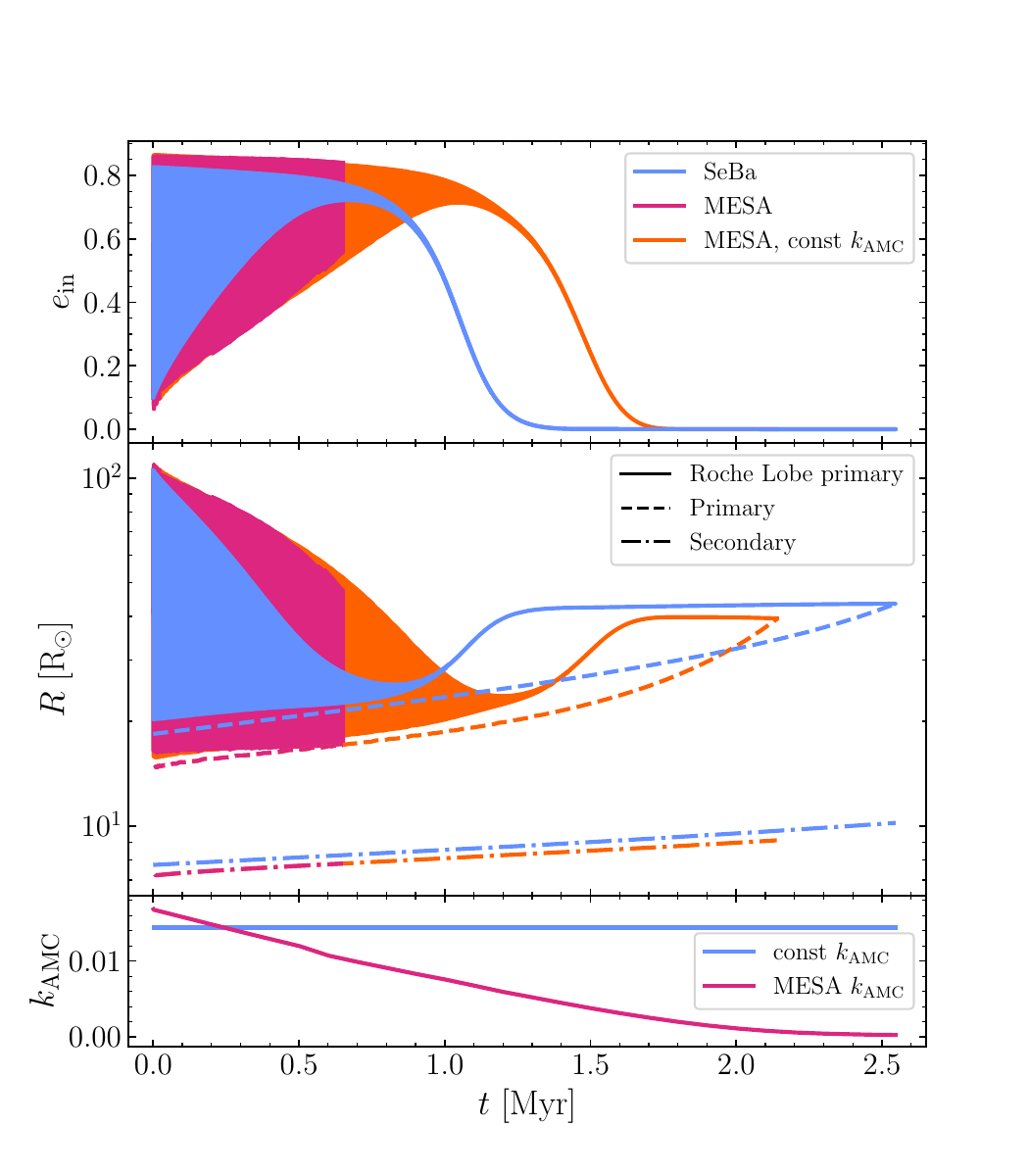}}
\caption{
Same as in Fig. \ref{eta_carina} with a larger inner separation (system labeled "$\eta$ Carinae*" in Table \ref{tab:initial_cond}). A model variation (\mesa with constant $k_{\rm AMC}$) is shown in orange.}
\label{eta_carina_star}
\end{figure}

In this appendix we present a triple system with initial conditions very similar to those of the system simulated in Sect. \ref{eta_car_section}, apart from the inner semi-major axis, which is slightly increased ($a_{\rm in}=1.1$\,au) compared to the original $\eta$ Carinae system ($a_{\rm in}=1$\,au). It corresponds to the system labeled "$\eta$ Carinae*" in Table \ref{tab:initial_cond}. Given the larger inner separation, the octupole term for this system is initially larger than for the original system ($\epsilon_{\rm oct}=0.052$), therefore the ZLK oscillations are more pronounced. We simulated this system with \seba and \mesaa, and also included a model variation where the stellar evolution is performed with \mesaa, but the apsidal motion constant is attributed the same fixed value $k_{\rm AMC}=0.0144$ as in the \seba simulation. The same quantities as in Fig. \ref{eta_carina} for this system are shown in Fig. \ref{eta_carina_star}, which illustrates the impact of the precession.
The \mesa model with a self-consistent value of $k_{\rm AMC}$ is subject to maximal ZLK oscillations, as the precession due to distortion are weaker compared to the precession due to the ZLK mechanism (for $t\ge 0.2$\,Myr, the apsidal motion constant of the primary retrieved from the \mesa structure gets smaller than the constant value used in the other simulations). The \mesa model variation shows an evolution more consistent with the system computed with \sebaa, even though differences due to the stellar evolution are still present.

Because the ZLK oscillations are larger in the default \mesa system, the primary star fills its Roche Lobe after only $\sim \!0.7$\,Myr, when the inner binary is still in an eccentric  orbit. The system in this case experiences an eccentric mass transfer episode. In contrast, ZLK oscillations in the \mesa model variation and the system simulated with \seba are quenched more efficiently due to a higher $k_{\rm AMC}$, and the systems in this case avoid RLOF during the eccentric phase. The systems experience mass transfer later ($\sim 2.1$\,Myr for the \mesa model variation and $\sim 2.5$\,Myr for the \seba model), when they have already circularized.

In Table \ref{tab:initial_cond}, the system of this section is not labeled as strictly divergent, as the nomenclature we followed is that of the paper by \citep{kum23} (in all cases an inner MT occurs), but given the above discussion, the evolutionary pathways of this system, as predicted by \seba and \mesaa, still differ.
\section{Apsidal motion constant evolution of the \mesa models}\label{AppC}
\begin{figure}[h]
\centering
\centerline{\includegraphics[trim=0.1cm 0cm 1cm 1.2cm, clip=true, width=1.02\columnwidth,angle=0]{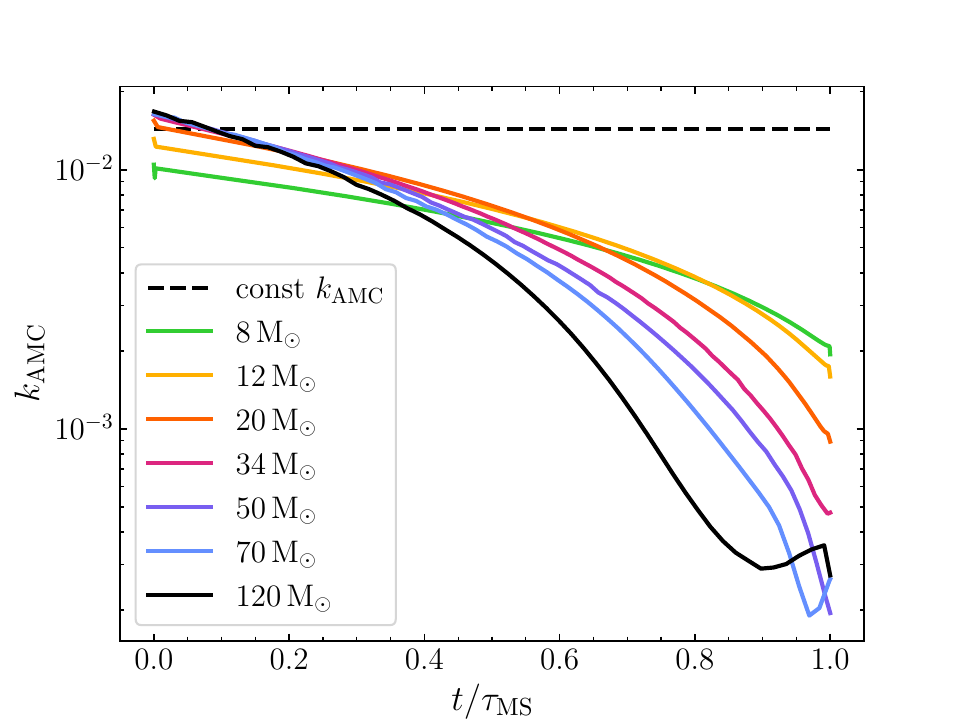}}
\caption{
MS evolution of the apsidal motion constant of the \mesa models for masses in the range $M\in[8,120]$\,M$_\odot$.}
\label{k2_time}
\end{figure}

The MS evolution of the apsidal motion constant $k_{\rm AMC}$ is represented in Fig. \ref{k2_time} for the whole range of masses of the grid ($M\in [8,120]$\,M$_\odot$), and with \texttt{Dutch\_scaling\_factor=0.333}. The constant MS value $k_{\rm AMC}=0.0144$ used in the simulations with \seba is shown as a reference. It can be noted that although the ZAMS value of the models is relatively consistent with the value used by default in \tress, the apsidal motion constant of the \mesa models decreases by more than one order of magnitude along the MS evolution. In terms of mass dependence, we note that lower mass models at ZAMS have lower $k_{\rm AMC}$ than higher mass ones, but a transition occurs around the middle of the MS and at TAMS the higher mass models have a lower $k_{\rm AMC}$, consistently with the models of \citep[e.g.,][]{cla04}. This can be explained by the fact that higher mass models have larger convective core, which results in a larger portion of the total mass being converted to helium at the TAMS and hence a higher density contrast (lower $k_{\rm AMC}$).
\section{\sci model of dynamical tides in eccentric orbits}\label{AppD}
In this appendix we provide a detailed description of the dynamical tide model used as an alternative to the \hu formalism in the simulations of Sect. \ref{ZLKTF}. The prescription we used is the Equation (9) in \sci, which consists of a complement to the \zah formalism. The prescription reads:

\begin{equation}
\scalebox{0.9}{$\begin{split}
&\restriction{\frac{\text{d}}{\text{d}t}\left(I\Omega_{\rm spin}\right)}{\rm Dyn}=\frac{3}{2}\frac{GM^2}{R}q^2 E_2\left(\frac{R}{a}\right)^6\cdot\Bigg\{s_{22}^{8/3}\text{sgn}\left(s_{22}\right)\\&+e^2\left(\frac{1}{4}s_{12}^{8/3}\text{sgn}\left(s_{12}\right)-5s_{22}^{8/3}\text{sgn}\left(s_{22}\right)+\frac{49}{4}s_{32}^{8/3}\text{sgn}\left(s_{32}\right)\right)\Bigg\}\\
&\restriction{\frac{\text{d}e}{\text{d}t}}{\rm Dyn}=-\frac{3}{4}e\left(\frac{GM}{R^3}\right)^{1/2}q(1+q)^{1/2}E_2\left(\frac{R}{a}\right)^{13/2}\cdot \\&\Bigg(\frac{3}{2}s_{10}^{8/3}\text{\rm sgn}\left(s_{10}\right)-\frac{1}{4}s_{12}^{8/3}\text{\rm sgn}\left(s_{12}\right)-s_{22}^{8/3}\text{\rm sgn}\left(s_{22}\right)+\frac{49}{4}s_{32}^{8/3}\text{\rm sgn}\left(s_{32}\right)\Bigg).
\end{split}
\label{zahn_corrected}$}
\end{equation}

Eq. \eqref{zahn_corrected} predicts the time evolution of the eccentricity and rotational velocity of a star subject to dynamical tides with radiative damping, i.e. gravity modes excited by the presence of a companion of mass $M_2 = qM$, where $M$ is the mass of the considered star and $q$ the mass ratio. We recall that the coefficients $s_{lm}$ are defined as:

\begin{equation}
    s_{lm} \equiv (l\Omega_{\rm orb} - m\Omega_{\rm spin})(R^3/GM)^{1/2}.
\end{equation}
When the two components of the binary system are massive enough to be subject to this type of tides (i.e. when they both possess a convective core and a radiative envelope), the equation can be applied to each component, and the evolution of the semi-major axis and eccentricity of the system can be obtained as $\dot a=\dot a_1 + \dot a_2$ and $\dot e=\dot e_1 + \dot e_2$, adding up the contributions of the effect of the dynamical tides on each component.

Imposing conservation of the total angular momentum\;$\mathcal{\dot L_{\rm tot}}= \mathcal{\dot L_{\rm spin}}+\mathcal{\dot L_{\rm orb}}$ and using Eq. \eqref{zahn_corrected}, the equation governing the evolution of the semi-major axis is:

\begin{equation}
\scalebox{0.9}{$\begin{split}
\restriction{\frac{\text{d}a}{\text{d}t}}{\rm Dyn}=-&3\left(\frac{GM}{R}\right)^{1/2}q(1+q)^{1/2}E_2\left(\frac{R}{a}\right)^{11/2}\cdot\\\Bigg\{\frac{3}{4}&s_{10}^{8/3}\text{sgn}\left(s_{10}\right)\cdot\frac{e^2}{1-e^2}\\+&s_{12}^{8/3}\text{sgn}\left(s_{12}\right)\cdot\left[\frac{1}{4}\frac{e^2}{\sqrt{1-e^2}}-\frac{1}{8}\frac{e^2}{1-e^2}\right]\\
+&s_{22}^{8/3}\text{sgn}\left(s_{22}\right)\cdot\left[\frac{1}{\sqrt{1-e^2}}\left(1-5e^2\right)-\frac{1}{2}\frac{e^2}{1-e^2}\right]\\+&s_{32}^{8/3}\text{sgn}\left(s_{32}\right)\cdot\left[\frac{49}{4}\frac{e^2}{\sqrt{1-e^2}}+\frac{49}{8}\frac{e^2}{1-e^2}\right]
\Bigg\}.
\end{split}
\label{semi_major}$}
\end{equation}
We implemented Equations \eqref{zahn_corrected} and \eqref{semi_major} in \tres and compared the evolutions of the orbital parameters of a few systems they provide to those obtained with the \hu prescription.

In order to interpret the results Eq. \eqref{zahn_corrected} and \eqref{semi_major} provide, it is useful to rewrite Eq. \eqref{zahn_corrected} as:

\begin{equation}
\scalebox{0.9}{$\begin{split}
\restriction{\frac{\text{d}}{\text{d}t}\left(I\Omega_{\rm spin}\right)}{\rm Dyn}&=\frac{3}{2}\frac{GM^2}{R}q^2 E_2\left(\frac{R}{a}\right)^6\cdot f_\Omega\left(\Omega_{\rm spin},\Omega_{\rm orb},e\right)\\
\restriction{\frac{\text{d}e}{\text{d}t}}{\rm Dyn}&=-\frac{3}{4}e\left(\frac{GM}{R^3}\right)^{1/2}q(1+q)^{1/2}E_2\left(\frac{R}{a}\right)^{13/2}\cdot f_e\left(\Omega_{\rm spin},\Omega_{\rm orb}\right),
\end{split}
\label{zahn_corrected_rewritten}$}
\end{equation}
where the functions $f_\Omega$ and $f_e$ are defined as:
\begin{equation}
\scalebox{0.8}{$\begin{split}
f_\Omega\left(\Omega_{\rm spin},\Omega_{\rm orb},e\right)&\equiv s_{22}^{8/3}\text{sgn}\left(s_{22}\right)+e^2\left(\frac{1}{4}s_{12}^{8/3}\text{sgn}\left(s_{12}\right)-5s_{22}^{8/3}\text{sgn}\left(s_{22}\right)+\frac{49}{4}s_{32}^{8/3}\text{sgn}\left(s_{32}\right)\right)\\
f_e\left(\Omega_{\rm spin},\Omega_{\rm orb}\right)&\equiv\frac{3}{2}s_{10}^{8/3}\text{\rm sgn}\left(s_{10}\right)-\frac{1}{4}s_{12}^{8/3}\text{\rm sgn}\left(s_{12}\right)-s_{22}^{8/3}\text{\rm sgn}\left(s_{22}\right)+\frac{49}{4}s_{32}^{8/3}\text{\rm sgn}\left(s_{32}\right).
\end{split}
\label{w}$}
\end{equation}
The signs of the functions $f_\Omega$ and $f_e$ are crucial as they determine whether the corresponding quantities ($\Omega$ and $e$) increase or decrease with time.

With this set of equations, pseudo-synchronization is reached for a given eccentricity when $f_\Omega\left(\Omega_{\rm spin},\Omega_{\rm orb},e\right)=0$. Because the equation governing the evolution of the angular velocity is directly obtained from \zah formalism without being combined to the \citep{hut81} formalism (as is is done following the \hu derivation), it provides different pseudo-synchronization values than what is obtained with the \hu prescription (where the pseudo-synchronization values are those predicted by \citealt{hut81}).

\begin{figure}[h]
\centering
\centerline{\includegraphics[trim=1.2cm 2.2cm 1.2cm 2.2cm, clip=true, width=1.02\columnwidth,angle=0]{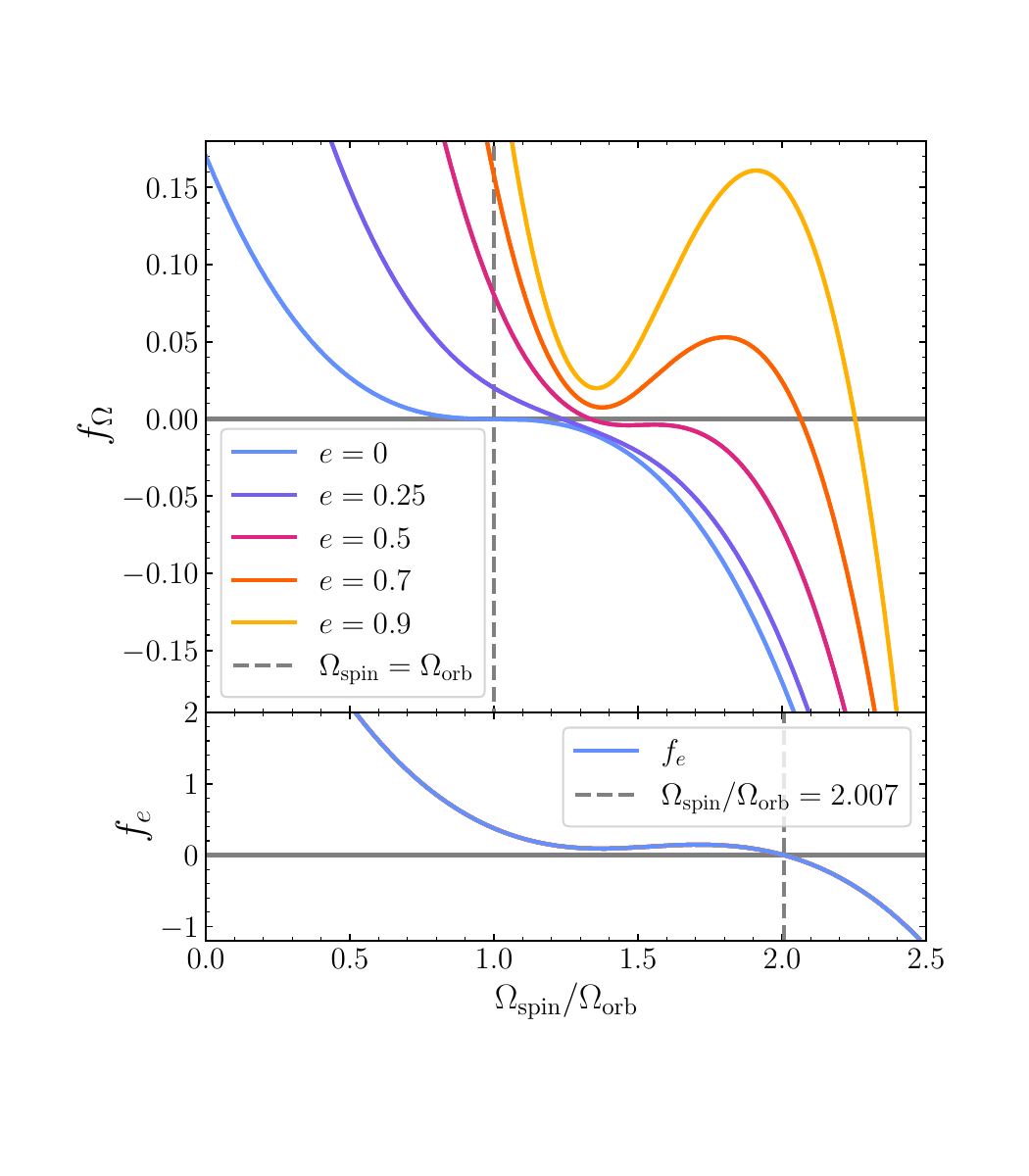}}
\caption{
\textit{Upper panel:} $f_\Omega$ as function of the ratio $\Omega_{\rm spin}/\Omega_{\rm orb}$ for different values of the eccentricity. The dashed gray line corresponds to $\Omega_{\rm spin}=\Omega_{\rm orb}$. \textit{Lower panel:} $f_e$ as function of the ratio $\Omega_{\rm spin}/\Omega_{\rm orb}$. The dashed gray line corresponds to the value of $\Omega_{\rm spin}/\Omega_{\rm orb}$ where $f_e$ changes sign.}
\label{f_omegas}
\end{figure}

In Fig. \ref{f_omegas}, $f_\Omega$ and $f_e$ are represented as functions of the ratio $\Omega_{\rm spin}/\Omega_{\rm orb}$ (for a star of star of mass $M=15$\,M$_\odot$, $R=4.7$\,R$_\odot$). As $f_\Omega$ also depends on the eccentricity, we show its shape for different values of the eccentricity ($e=[0,0.25,0.5,0.7,0.9]$). Although the functions $f_\Omega$ and $f_e$ depend on both $\Omega_{\rm spin}$ and $\Omega_{\rm orb}$, their zeros only depend on the ratio $\Omega_{\rm spin}/\Omega_{\rm orb}$. For this reason we only show the functions for a fixed period of $P=1.2$\,days, corresponding to $\Omega_{\rm orb}=6.06\times 10^{-5}$\,rad/s.
Among the important features in the shape of $f_\Omega$ and $f_e$, we can note the following:
\begin{enumerate}
    \item The zeros of $f_\Omega$ (i.e. the values of the ratio of $\Omega_{\rm spin}/\Omega_{\rm orb}$ for which the star is pseudo-synchronized) increase with the eccentricity, which is the expected behavior for pseudo-synchronization. For circular orbits, the zero of $f_\Omega$ is found when $\Omega_{\rm spin}/\Omega_{\rm orb}=1$ as expected.
    \item The function $f_e$ is not positive everywhere. Indeed, it changes sign at $\Omega_{\rm spin}/\Omega_{\rm orb}\approx 2.007$. This result is independent of the eccentricity as $f_e$ only depends on $\Omega_{\rm spin}$ and $\Omega_{\rm orb}$ and was already obtained by \zah (see Fig. 2 in the paper). It implies that, for a star rotating above this threshold, the dynamical tides makes the eccentricity increase. This goes against the standard picture that tides always act towards the circularization of the system, but directly stems from \zah formalism.
    \item One can note looking at the curves of $f_\Omega$ that for systems with high eccentricities ($e\gtrsim 0.7$), the pseudo-synchronization value is larger than the threshold value of $\Omega_{\rm spin}/\Omega_{\rm orb}=2.007$ at which $f_e$ changes sign. This implies that for any pseudo-synchronized star in a system with $e\gtrsim 0.7$, the dynamical tides makes the eccentricity of the system increase. 
\end{enumerate}
It is useful to obtain the value of the ratio $\Omega_{\rm spin}/\Omega_{\rm orb}$ for which the star is pseudo-synchronized as a function of the eccentricity. We can thus define the corresponding function of the eccentricity $f_{\rm sync}(e)$ in the following way:
\begin{equation}
    f_{\rm sync}(e)\equiv\max\restriction{\frac{\Omega_{\rm spin}}{\Omega_{\rm orb}}}{f_\Omega\left(\Omega_{\rm spin},\Omega_{\rm orb},e\right)=0}.
\end{equation}
The maximum notation is used as there is a small interval of eccentricities for which $f_\Omega$ has several zeros. Indeed, as can be seen looking at Fig. \ref{f_omegas}, the graph of $f_\Omega$ is similar to that of a polynomial function of degree 3, which may have one or three zeros. It turns out that for eccentricities in the interval $e\in[0.524, 0.566]$, the function $f_\Omega$ has three zeros. In order to correctly define the function $f_{\rm sync}$ for any eccentricity, we take the maximum of the three values in this interval.

Because the dependence in $R/a$ is larger in the the eccentricity equation than in the angular velocity equation, pseudo-synchronization usually occurs faster than circularization. It is therefore useful to determine whether the action of the tides on a pseudo-synchronized star will tend to make the eccentricity increase or decrease. To this purpose, we define the function $f_{\rm circ}(e,\Omega_{\rm orb})$ as:
\begin{equation}
    f_{\rm circ}(e,\Omega_{\rm orb})\equiv\restriction{f_e(\Omega_{\rm spin},\Omega_{\rm orb})}{\frac{\Omega_{\rm spin}}{\Omega_{\rm orb}}=f_{\rm sync}(e)}.
\end{equation}
$f_{\rm circ}$ corresponds to the value of the function $f_e$ for any pseudo-synchronized star. It is eccentricity dependent as the value of the ratio $\Omega_{\rm sync}/\Omega_{\rm orb}$ for which pseudo-synchronization is reached depends on the eccentricity. $f_{\rm circ}$ still depends on $\Omega_{\rm orb}$ (or $\Omega_{\rm spin}$) as the pseudo-synchronization condition only constrains the ratio $\Omega_{\rm sync}/\Omega_{\rm orb}$, but does not constrain both $\Omega_{\rm sync}$ and $\Omega_{\rm orb}$. Since the zeros of both $f_e$ and $f_\Omega$ only depend on the ratio $\Omega_{\rm sync}/\Omega_{\rm orb}$, it can be shown that the zero of $f_{\rm circ}$ is the same for any value of $\Omega_{\rm orb}$, and is reached for a single value of the eccentricity of $e\approx 0.659$. 

\begin{figure}[h]
\centering
\centerline{\includegraphics[trim=1.8cm 1cm 1.4cm 1.2cm, clip=true, width=1.02\columnwidth,angle=0]{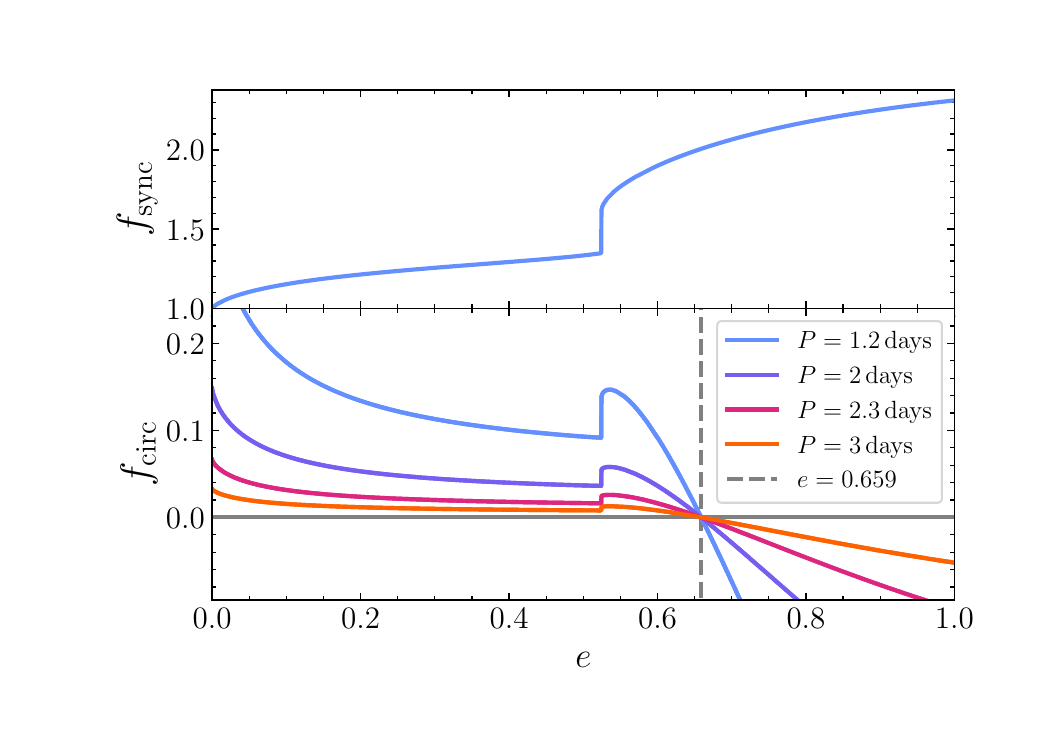}}
\caption{
\textit{Upper panel:} $f_{\rm sync}$ as function of the eccentricity. \textit{Lower panel:} $f_{\rm circ}$ as function of the eccentricity for different values of $\Omega_{\rm orb}$. The dashed gray line corresponds to the value of the eccentricity where $f_{\rm circ}$ changes sign.}
\label{f_e}
\end{figure}

We show in Fig. \ref{f_e} $f_{\rm sync}$ and $f_{\rm circ}$ as functions of the eccentricity. We represent $f_{\rm circ}$ for different values of the orbital angular velocity.

The following comments can be made regarding Fig. \ref{f_e}:
\begin{enumerate}
    \item The visible discontinuity of $f_{\rm sync}$ at $e\approx 0.566$ can be imparted to the fact that in the interval $e\in [0.524,0.566]$, $f_\Omega$ possesses three pseudo-synchronization values, and we take the maximum of the three. $f_{\rm sync}$ shows an expected behavior: it increases with the eccentricity and reaches unity at $e=0$, which is interpreted as standard synchronization.
    \item 
    $f_{\rm circ}$ also presents a discontinuity at $e\approx 0.566$, which is not surprising given that it is defined with respect to $f_{\rm sync}$.
    \item The most striking feature of $f_{\rm circ}$ is that it changes sign at $e\approx 0.659$. The interpretation of this result is that according to this model, dynamical tides tend to increase the eccentricity of any pseudo-synchronized star in an orbit with $e\gtrsim 0.659$. This can in practice never make the eccentricity reach unity as the increase of $e$ also implies an increase of $a$ by conservation of angular momentum, which will at some point quench the effect of the tides, given the strong dependence in $a$ in the eccentricity evolution.
    \item Given this interpretation of $f_{\rm circ}$, we can understand that the tidal evolution of systems in highly eccentric orbits as predicted by \sci may differ significantly from what is predicted following the \hu formalism. In the case of triple systems where the inner binary is subject to large eccentric ZLK oscillations, where the inner eccentricity can reach extreme values, the dynamical tides in this case will not tend to make the eccentricity decrease but increase even more, which can have important consequences for the evolution of the system (e.g. RLOF in an eccentric orbit).
\end{enumerate}
\end{appendix}
\end{document}